\documentclass[a4paper,11pt]{article}
\pdfoutput=1 % if your are submitting a pdflatex (i.e. if you have
\usepackage{jheppub, bm, amscd} % for details on the use of the package, please
\usepackage[T1]{fontenc} % if needed

\newcommand{\met}{\not \!\! E_T}

\title{\boldmath Exploring Theory Space with Monte Carlo Reweighting}

\author[a,1]{James S. Gainer,}
\author[b]{Joseph Lykken,} 
\author[a]{Konstantin T. Matchev,}
\author[c]{Stephen Mrenna,}
\author[d,1]{and Myeonghun Park\note{Corresponding author.}}

\affiliation[a]{Physics Department, University of Florida, P. O. Box
  118440, Gainesville, FL 32611, USA}
\affiliation[b]{Theoretical Physics Department, Fermilab, P. O. Box
  500, Batavia, IL 60510, USA}
\affiliation[c]{SSE Group, Computing Division, Fermilab, P. O. Box
  500, Batavia, IL 60510, USA}
\affiliation[d]{Kavli Institute for the Physics and Mathematics of the
  Universe (WPI), Todai Institutes for Advanced Study, The University
  of Tokyo, 5-1-5 Kashiwanoha, Kashiwa, 277-8583, Japan }

\emailAdd{gainer@phys.ufl.edu}
\emailAdd{lykken@fnal.gov}
\emailAdd{matchev@phys.ufl.edu}
\emailAdd{mrenna@fnal.gov}
\emailAdd{myeonghun.park@ipmu.jp}

\abstract{Theories of new physics often involve a large number of
  unknown parameters which need to be scanned.  Additionally, 
  a putative signal in a particular channel may be due to a 
  variety of distinct models of new physics.
  This makes experimental attempts to
  constrain the parameter space of motivated new physics models
  with a high degree of generality quite challenging.
  We describe how the reweighting of events may allow this challenge to be
  met, as fully simulated Monte Carlo samples generated for arbitrary benchmark
  models can be effectively re-used.  In particular, we suggest
  procedures that allow more efficient collaboration between theorists
  and experimentalists in exploring large theory parameter spaces in a
  rigorous way at the LHC.}

\preprint{IPMU 14-0100}
\begin{document} 
\maketitle
\flushbottom
%\linenumbers
\section{Introduction}
\label{sec:intro}

Monte Carlo (MC) simulation is an essential and ubiquitous tool in particle
physics~\cite{Cowan, Nason, Beringer:1900zz, Buckley:2011ms}.
For realistic studies, such simulation generally
includes the modeling of the
detector response as well as the underlying physical process.  While
theoretical studies generally require only ``fast'' detector
simulation, such as that performed by {\tt PGS}~\cite{PGS4} or
{\tt Delphes}~\cite{deFavereau:2013fsa},
the modeling of the detector response to a given physics event at the
experimental level, using frameworks such as 
{\tt GEANT}~\cite{Agostinelli:2002hh}, is necessarily a
time-consuming and computationally expensive process.

The challenge of accommodating the current lack of direct evidence
for new physics within frameworks that address the limitations of the
Standard Model (SM) leads to models with many parameters, such as the
general gauge mediation~\cite{Giudice:1998bp, Carpenter:2008wi,
  Intriligator:2008fr, Meade:2008wd, Carpenter:2008he, Buican:2008ws, Craig:2013wga} and
pMSSM~\cite{Berger:2008cq, Cotta:2009zu, AbdusSalam:2009qd, Sekmen:2011cz,
  Arbey:2011un, CMS:2013rda} frameworks in low-energy Supersymmetry
(SUSY)\cite{Martin:1997ns, Drees:2004jm, Baer:2006rs, Dine:2007zp}.  
Even when no particularly compelling model of Beyond
the SM (BSM) physics exists for a given process, it may be useful to
parameterize potential deviations from the SM using an effective
theory~\cite{Wilson:1973jj, Georgi:1990um, Georgi:1994qn,
  Kaplan:1995uv, Manohar:1996cq, D'Ambrosio:2002ex, Rothstein:2003mp,
  Rajaraman:2011wf},
which may involve many operators and hence many
parameters.

In addition to the situations above which involve many continuous
parameters, there are also issues with discrete parameters; in
particular there can be many choices for the spin assignments for
particles in a given event topology.  A well-known example of this is
the fact that most if not all SUSY topologies are also produced
in models of Universal Extra Dimensions
(UED)~\cite{Appelquist:2000nn,Cheng:2002ab}.  It is therefore
desirable to study, at the very least, the UED counterparts of SUSY
signal benchmark points.\footnote{Ideally one would like to avoid
theoretical prejudice completely, by going beyond SUSY
and UED spin assignments.}  

Clearly experimentalists at the CERN Large Hadron Collider (LHC) and
similar experiments are faced with an overwhelming multitude of
theoretical models which could be searched for. To determine
the properties of a given signal in the detector generally requires
detector-level MC simulation (fullsim) of that signal, e.g., with {\tt GEANT}, 
which is a time-consuming process, as noted above.  
To repeat this procedure for the myriad possibilities
suggested by theorists is simply not possible.

Attempts have been made to solve this
problem~\cite{Cranmer:2010hk, Rao:2012sx, Drees:2013wra, Papucci:2014rja}, 
perhaps most notably
by encouraging theorists to test the predictions of their models against
suitably chosen experimental results (e.g., limits on cross-sections times 
branching fractions as a function of the mass spectrum). 
However, an actual discovery will still require the use of fullsim samples to obtain
realistic distributions for the kinematic variables of interest (such as test statistics). 

We propose and describe the following procedure for the experimental
analysis of a large number of signal hypotheses:
\begin{enumerate}
\item Generate a large set of unweighted sample events at generator
  level, $\{G_i\}$ for a specific model and point in that model's
  parameter space, which we term $(A,\bm{\alpha})$ (where $A$
  describes the model and $\bm{\alpha}$ specifies a point in that
  model's parameter space); a reasonable choice
for the model, $A$, would be a simplified
model~\cite{ArkaniHamed:2007fw,Alwall:2008ag,Alves:2011wf,ATLAS:2011lka,
Okawa:2011xg,ATLAS:2011ffa,Chatrchyan:2013sza}.
\item Pass the generator-level events, $\{G_i\}$ through realistic detector
  simulation.  Select desired events (``apply cuts'') to obtain a set
  of detector level events $\{D_i\}$.  (Every generator-level event
  $G_i$ maps to some detector-level event, $D_i$, but not every
  detector-level event passes our cuts.) 
\item Note that the cross section after acceptances and efficiencies,
  as well as the detector-level predictions for all distributions in
  signal hypothesis $(A,\bm{\alpha})$ can be obtained from the
  resulting event sample.
\item Choose some other model, $B$, and parameter space point,
  $\bm{\beta}$.  Assign each detector-level event $D_i$ a weight 
$w(B,\bm{\beta},G_i)/ w(A,\bm{\alpha},G_i)$, which \emph{depends only
  on generator-level information} and uses, in particular, the
``truth'' values of particles such as neutrinos or neutralinos which
are not observed in the detector.
\end{enumerate}
Because the reweighting involves only generator-level information, it
is very fast compared with detector simulation.  As we will show in
the following section, where we derive and justify the method, this is
not an approximation-- in the limit of infinite MC samples, this
method reproduces all distributions and quantities of interest
exactly.  

Therefore, a relatively efficient method for examining the large
``theory'' space would be for experimentalists to provide theorists
with the hard-process/ parton-level events which they have subjected
to detector simulation.  The theorists can then calculate the weights
of these events under the original model/ parameter space point as well
as the weights of these events for any desired point in model/
parameter space.  The experimentalists can take these weights and
obtain any desired distribution \emph{at detector level} from
reweighting their detector-level events.   (Of course the ``theorist''
in this description could also be an experimentalist interested in the
model being studied.)  
While, as we will discuss
below, practical issues \emph{may} arise for finite MC samples, in
general this method could vastly extend the reach into theory space of
LHC analyses.  

We reiterate that the MC events being shared here are ``truth''
events.  The momenta of all ``invisible'' final state particles (such
as neutrinos or neutralinos) are fully specified.  Therefore the
calculation of weights in different scenarios is numerically trivial;
the time-consuming integration over undetermined momenta which is
necessary in the Matrix Element Method~\cite{kondo1, kondo2,
  kondo3, dalitz, oai:arXiv.org:hep-ex/9808029, vigil, canelli, abazov,
  Gainer:2013iya},  and which can be
performed by tools such as MadWeight~\cite{Artoisenet:2008zz,
  Artoisenet:2010cn} is not necessary.

In Section~\ref{sec:mc}, we will provide a brief overview of MC
simulation in particle physics.  It is not our intent to provide a
review (see, e.g., Refs.~\cite{Nason, Buckley:2011ms}).  We wish only
to remind the reader that a set of unweighted hard-process
parton-level events $\bm{x}_{i,{\rm true}}$ can be used to generate a
set of unweighted events after showering, hadronization, detector
simulation, object reconstruction, etc., which are given by
\begin{equation}\label{eq:T}
\bm{x}_{i,{\rm objects}} = T(\bm{x}_{i,{\rm objects}},
  \bm{x}_{i,{\rm true}}) \bm{x}_{i,{\rm true}}.
\end{equation}
The important points about the generalized transfer function $T$,
which incorporates the vast amount of physics briefly summarized in
Section~\ref{sec:mc}, are
\begin{enumerate}
  \item It does not depend on the new physics model or parameters.
  \item It is unitary.  Every hard-process event
    simulated corresponds to a detector-simulated event with objects
    reconstructed, even if, e.g., due to the failure to reconstruct
    the physics objects corresponding to some partons, it will not be
    included in our calculations of distributions for the relevant
    final state.
    \item As $T$ is unitary, if we take a set of unweighted events
      $\bm{x}_{i,{\rm true}}$ and simulate each event, $\bm{x}_{i,{\rm
          true}}$, or, in equivalent language, choose a random value of $\bm{x}_{\rm
        objects}$ according to the probability distribution function
      (PDF), $T(\bm{x}_{i,{\rm objects}},
      \bm{x}_{i,{\rm true}})$ for each $\bm{x}_{i,{\rm true}}$, then
      we will obtain an unweighted set of detector-simulated and
      object-reconstructed events $\{\bm{x}_{i,\rm{objects}}\}$.
\end{enumerate}
If the reader is willing to accept these facts about $T$, then
Section~\ref{sec:mc} may be safely skipped.     

Reweighting itself is described in Section~\ref{sec:reweighting}.
To demonstrate its use, we then provide a few examples of reweighting in action.
In Section~\ref{sec:h4l} we study a signal model with a multi-dimensional parameter space, 
while in Section~\ref{sec:antler} we use reweighting to analyze angular
correlations needed to determine the spins of the new particles.  
In Section~\ref{sec:clustering} we demonstrate that reweighting works even in the 
presence of showering, hadronization, jet clustering, and detector simulation. 
We present some useful notes for practitioners of our method 
in Section~\ref{sec:practical}. Section~\ref{sec:conclusions} is reserved for our conclusions. 

\section{Monte Carlo Overview}
\label{sec:mc}

\subsection{Monte Carlo Basics}
\label{subsec:basics}

The integral $F$ of an arbitrary function $f(x)$ over some interval
$(x_-,x_+)$ is:
\begin{equation}
F \equiv F(x_{+}) - F(x_{-}) = \int_{x_{-}}^{x_{+}} dx f(x) = \langle f \rangle \int_{x_{-}}^{x_{+}} dx
= \langle f \rangle (x_{+} - x_{-}),
\end{equation}
where $F(x)$ is the antiderivative of $f(x)$ and $\langle f \rangle$ is
the average value of $f(x)$.
The Monte Carlo technique for estimating $\langle f \rangle$ is 
to select $N$ values of $x_i = x_{-} + r_i (x_{+} - x_{-})$,
where $r_i$ is a uniform random variate in the range (0,1),
and take the average of $f(x)$ at these points:
\begin{equation}
\langle f \rangle = \frac{1}{N} \sum_{i=0}^{i=N-1} f(x_i).
\label{faverage}
\end{equation}
The points $x_i$ used to evaluate this sum can be viewed as a set of ``events" 
with weights $f(x_i)$. 

To compute the MC average of a different function, $g(x)$, we similarly have:
\begin{equation}
\langle g \rangle = \frac{1}{N} \sum_{i=0}^{i=N-1} g(x_i)
= \frac{1}{N} \sum_{i=0}^{i=N-1} f(x_i)\, \frac{g(x_i)}{f(x_i)}.
\label{gaverage}
\end{equation}

Importance sampling considers a change in the measure of sampling in
order to reduce the variance of the MC estimate:
\begin{equation}
\int_{x_{-}}^{x_{+}} dx f(x) = \int_{x_{-}}^{x_{+}} dx f(x) \frac{g(x)}{g(x)}= 
\int_{x_{-}}^{x_{+}} dx g(x) \frac{f(x)}{g(x)}\equiv
\left\langle \frac{f}{g} \right\rangle_{g} \int_{x_{-}}^{x_{+}} g(x) dx.
\label{impsamp}
\end{equation}
In many applications, $g(x)$ can be integrated analytically and has an
antiderivative $G$ with a inverse $G^{-1}$.  Then $x_i = G^{-1}(r_i +
G(x_{-}))$.
From (\ref{faverage}) and (\ref{impsamp}) it follows that:
\begin{equation}
\left\langle \frac{f}{g} \right\rangle_{g} = \frac{1}{N} \sum_{i=0}^{i=N-1} 
\left.\frac{f(x_i)}{g(x_i)}\right\vert_{x_i = G^{-1}(r_i + G(x_{-}))},
\end{equation}
where we emphasize that the average is with respect to the function $g$.
It is not necessary to have a closed form for $g$, $G$ or $G^{-1}$, and
this will be the case for the practical application of these formulas discussed later.
In fact, the
selection of integration points from $g(x)$ may be performed numerically
using, for example, {\tt VEGAS} \cite{Lepage:1977sw}.  As a trivial example,
assume a set of $N$ unweighted events generated according to $g(x)$, so that
each event has a weight $\sigma/N$.   Then, choosing $f(x)=g(x)$, we have:
\begin{equation}
\langle g \rangle = \left\langle 1 \right\rangle  \int_{x_{-}}^{x_{+}} dx g(x)
\simeq \sum_{i=0}^{N-1} \frac{\sigma}{N} = \sigma.
\end{equation}
Even though the weights are uniform in this case, the values of $x_i$ associated
with each event are distributed according to $g(x)$ and can be used to
calculate averages or construct histograms of other quantities based
on $x$.

\subsection{Parton-Level Event Generation}

In the applications discussed here, we are interested in performing a
reweighting of events at the parton-level.   
The parton-level is usually the first non-trivial level of
approximation where physics beyond the Standard Model is needed.  
In the examples considered here, we focus on hadron-hadron collisions.  
The basic quantities of interest are the kinematic properties of
events involving some new particles or interactions.   
We limit ourselves to processes of the type 
$1+2 \to i+j+\cdots+n$.  Then, the partonic cross section is proportional to
\begin{equation}
f_{1/h}(x_1,Q^2) f_{2/h}(x_2,Q^2) \bigg\vert {\cal A}[1,2 \to i,j,\cdots,n]  
\bigg\vert^2,
\end{equation}
where the amplitude for this process is $\cal A$.
Imagine now, another process with the {\it exact} same initial and final state,
but described by a different amplitude ${\cal A^\prime}$.
The relative probability of this process with respect to the former is given
simply by $|{\cal A^\prime}|^2/|{\cal A}|^2$.
This parton-level description is often called the ``hard-process''
and describes the physics occurring
at energies $Q \gg \Lambda_{\rm QCD}$, or equivalently length/ time
scales $\ll 1/\Lambda_{\rm QCD}$.  
The expression for the partonic cross section can be cast in the
form of a probability distribution that generates parton-level ``events'' or
configurations of kinematic variables $x_i$.
(In this heuristic description, we can think of e.g. the helicity and
color of all final state partons as specified; in general, we need
to generate all relevant final state helicities and color structures.)
These events might be the output of a program such as
{\tt MadGraph5}~\cite{Alwall:2011uj}, {\tt
  CalcHEP}~\cite{Belyaev:2012qa}, or {\tt
  CompHEP}~\cite{Pukhov:1999gg, Boos:2004kh}; these programs
can also provide code that can be used later in reweighting.

Since our interest is in the
reweighting of new physics signal events, the hard-processes we
consider will involve some new physics model, $A$, described by some
parameters $\bm{\alpha}$.  We will term this ``theory point''
$(A,\bm{\alpha})$.
Our goal is to obtain all cross sections
and distributions for a different model $B$, for the parameter point,
$\bm{\beta}$.  Of course, it may be that we are scanning the parameter
space of model $A$, in which case $B=A$ and we are considering
parameter points $\bm{\alpha}_1, \bm{\alpha}_2, ...$.  Also, for
some processes, like Higgs boson production and decay, it may make sense for
$(A,\bm{\alpha})$ to be the SM rather than a new physics model; 
we will then reweight to obtain distributions for the non-SM theory
point $(B, \bm{\beta})$.

\subsection{Particle-Level Event Generation}

For now we focus on perhaps the primary use of Monte Carlo event generators,
which is to
produce a sample of unweighted events of particles to be input
to a realistic detector simulation.
Generating events is important, because we are not interested only
in the cross section, subject to arbitrary event selection (``cuts''),

\begin{equation}\label{eq:sigma}
\sigma_G = \int_{\mathcal{D}_G} \frac{d \sigma(A,\bm{\alpha};
  \bm{x}_{\rm true})}{d\bm{x}_{\rm true}} d\bm{x}_{\rm true},
\end{equation}
but also in the kinematic distribution, $\rho(V)$, for a kinematic variable of interest,
$V(\bm{x}_{\rm true})$, subject to certain cuts:

\begin{equation}\label{eq:rho}
\rho(V^\prime) = \frac{1}{\sigma_G} \int_{\mathcal{D}_G} \frac{d
  \sigma(A,\bm{\alpha};
  \bm{x}_{\rm true})}{d\bm{x}_{\rm true}} 
\delta(V^\prime - V(\bm{x}_{\rm true}))
d\bm{x}_{\rm true}.
\end{equation}
Here $\mathcal{D}_G$ is the space of events ($\bm{x}_i$) which pass
cuts.  These are not the final, detector-level cuts.  However, cuts
are often applied at this stage, either to provide a cut-off for infrared
divergent quantities or simply to reduce the time spent on
detector-simulation of events that will not pass the triggers or the final,
detector-level cuts.
If we generate a sample of $N$ unweighted events
$\{\bm{x}_i\}$ in the allowed region, and, in the course of generating the events
determine the total cross section for the specified process to be
$\sigma_{G, {\rm MC}}$, then, of course, we can approximate the cross section after cuts by
$\sigma_G \approx \sigma_{G, {\rm MC}}$, while the normalized distribution of an
arbitrary kinematic variable, $\rho$ (as in Eq.~(\ref{eq:rho})) can be
approximated by forming the histogram 

\begin{equation}\label{eq:rho-mc}
  \int_{V_{\rm min}}^{V_{\rm max}} \rho(V) \, dV \approx \frac{1}{N} \sum_i \left\{
  \theta(V(\bm{x}_i) - V_{\rm min}) - \theta(  V(\bm{x}_i) - V_{\rm max} ) \right\},
\end{equation}
where $(V_{\rm min},V_{\rm max})$ is a bin interval.
Note that $\theta(V(\bm{x}_i) - V_{\rm min}) - \theta( V(\bm{x}_i) - V_{\rm max})$ 
is $1$ if $V_{\rm min} < V(\bm{x}_i) < V_{\rm max}$ and
$0$ otherwise.  The generalization of Eq.~(\ref{eq:rho}) and
Eq.~(\ref{eq:rho-mc}) to multidimensional observables is obvious.

The cross section in Eq.~(\ref{eq:sigma}) and the distributions
$\rho(V)$ in Eq.~(\ref{eq:rho}) and
Eq.~(\ref{eq:rho-mc}) have been calculated considering only the
physics from the hard process.  To obtain accurate distributions for
quantities measured in the detector, we must utilize
general purpose MC generators and/or showering tools like 
{\tt ARIADNE}\cite{Lonnblad:1992tz}, {\tt HERWIG}\cite{Corcella:2000bw},
{\tt HERWIG++}\cite{Bahr:2008pv}, {\tt PYTHIA
  6}\cite{Sjostrand:2006za}, {\tt PYTHIA 8}\cite{Sjostrand:2007gs},
{\tt SHERPA}\cite{Gleisberg:2003xi}, or {\tt
  VINCIA}\cite{Skands:2006ac, Larkoski:2013yi, Giele:2013ema}.  With
these tools we must simulate initial and final state radiation (ISR and
FSR), decay resonances with sufficiently short life times, and hadronize
colored objects.  While modeling the hadronic physics we must also
consider, e.g., the physics of the underlying
event\cite{Field:2010bc, Field:2011iq, Field:2012jv, Field:2012kd,
  Skands:2013asa}.
 
Formally, we can think of events at each level of the simulation as
living in different spaces.  Thus the initial, unweighted,
hard-process parton-level event undergoes a transformation:

\begin{eqnarray}
\begin{CD}
\bm{x}_{i, {\rm true}} @>>{S:~\rm showering}> \bm{x}_{i, {\rm
    showered}} @>>{H:~\rm hadronization}> \bm{x}_{i, {\rm hadron}}
\end{CD}.
\end{eqnarray}
We note that the dimensions of each $\bm{x}_i$ vector are in general
quite different.  Showering adds particles to the event via ISR and
FSR, and decays replace
resonances with two or more daughter particles.  Also, colored partons,
either before or after showering and decays, are obviously not 
in one-to-one correspondence with final state hadrons.  

The mappings $S$ and $H$ are unitary (provided we do not impose
additional cuts at this stage).  
In the case of these QCD processes, this ability to separate physics
at different length scales (and to consider the composition of
probabilities rather than amplitudes) results both from factorization
and from
more specific results like the KLN theorem~\cite{Kinoshita:1962ur, Lee:1964is} (see e.g.,
Refs.~\cite{Nason} and~\cite{Albino:2008gy} and references therein for more discussion
of these points). 
Thus, for a particular event $\bm{x}_{i,\rm{true}}$,

\begin{equation}
\int d \bm{x}_{\rm showered} S(\bm{x}_{\rm showered}, \bm{x}_{i,\rm{true}}) = 1.
\end{equation}
Likewise, for some particular showered event
$\bm{x}_{i,~\rm{showered}}$ we would have,

\begin{equation}
\int d \bm{x}_{\rm hadron} H(\bm{x}_{\rm hadron}, \bm{x}_{i,\rm showered}) = 1.
\end{equation}
Obviously, this implies that

\begin{equation}
\int d \bm{x}_{\rm showered} d\bm{x}_{\rm hadron} H(\bm{x}_{\rm hadron}, \bm{x}_{\rm showered}) 
S(\bm{x}_{\rm showered}, \bm{x}_{i,\rm{true}} )
= 1.
\end{equation}
It is clear that if we start with a set of unweighted hard-process
generator-level events
$\{\bm{x}_{i,{\rm true}}\}$, and then, for each $\bm{x}_{i,{\rm
    true}}$ choose a random value for
$\bm{x}_{i,\rm showered}$ according to the PDF $S(\bm{x}_{\rm
  showered}, \bm{x}_{i,\rm{true}})$
corresponding to $\bm{x}_{i,\rm{true}}$, then
we will obtain a set of unweighted, showered events $\{\bm{x}_{i,{\rm
    showered}}\}$.  We can obviously continue this procedure by
  choosing, $\bm{x}_{i,\rm{hadron}}$ according the the PDFs
  $H(\bm{x}_{\rm hadron}, \bm{x}_{i,\rm{showered}})$, thereby
  obtaining a set of unweighted, hadron-level events $\{\bm{x}_{i,{\rm
      hadron}}\}$.  That we perform this random selection according to
  the PDFs $S(\bm{x}_{\rm showered}, \bm{x}_{i,\rm{true}})$ and
  $H(\bm{x}_{\rm hadron}, \bm{x}_{i,\rm{showered}})$ using a
  simulation program such as {\tt HERWIG} or {\tt PYTHIA} does not
  affect the argument.
  We remind the reader that only the generation of the initial set of
  hard-process parton-level events $\{\bm{x}_{i,\rm{true}}\}$ had
  anything to do with the new physics theory point $(A,
  \bm{\alpha})$.  The subsequent showering and hadronization depends
  only on the hard process event $\bm{x}_{\rm true}$ (by which we
  include both the four momenta of the particles and the particular
  color structure which was generated) and SM parameters, such as
  $\alpha_S(Q)$.

\subsection{Detector-level Monte Carlo Simulation and Reconstruction}

In this subsection we briefly describe the process of performing detector-level MC
simulation, in order to explain why reweighting
detector-level events based on the hard-process matrix elements works.
  Two further stages of event evolution,
  detector simulation and object reconstruction,
  also decompose into the
  product of PDFs: 

  \begin{eqnarray}
    \begin{CD}
      \bm{x}_{i, {\rm hadron}} @>>{D:~\rm detector~simulation}> \bm{x}_{i, {\rm
          detector}} @>>{R:~\rm object~reconstruction}> \bm{x}_{i, {\rm objects}}
    \end{CD}.
  \end{eqnarray}
Detector simulation refers to determining the tracks produced and
energy deposited in various parts of the detector, using {\tt GEANT}.
So in general $\bm{x}_{i, {\rm detector}}$ does not describe the
four-momenta of particles, but the properties of tracks, energy in
calorimeter cells or towers, etc.  In order to do physics, we must
take this raw detector output and map it into ``physics objects'',
such as ``electrons'', ``muons'', or ``jets''.  The reconstruction of
jets, of course, is especially non-trivial; significant work has gone
into the theoretical and experimental understanding of jet
algorithms~\cite{Ellis:2007ib, Salam:2009jx}.  For our purposes, it suffices to note that
the standard simulation procedures are formally analogous to the showering and
hadronization described in that they can be viewed as obtaining
 a set of unweighted
detector-level events $\bm{x}_{i, {\rm detector}}$ corresponding to
hadron-level event $\bm{x}_{i, {\rm hadron}}$, by choosing a
random value of $\bm{x}_{i, {\rm detector}}$ from the PDF
$D(\bm{x}_{\rm detector}, \bm{x}_{i, {\rm hadron}})$ and obtaining an
event consisting of reconstructed physics objects, $\bm{x}_{i,\rm
  objects}$ from this detector event, $\bm{x}_{i,\rm detector}$ using
the PDF $R(\bm{x}_{\rm objects}, \bm{x}_{i,\rm detector})$.

In fact, we can compose the four PDFs, $S$, $H$, $D$, and $R$ into the
generalized-transfer function $T$ from Eq.~\ref{eq:T}, which, as a
product of normalized PDFs (as each step is unitary),
is also a normalized PDF.  While we did not discuss MC
tuning~\cite{Waugh:2006ip, Skands:2009zm, Buckley:2009bj,
  Skands:2010ak}, 
or pileup effects explicitly, we note that
these effects contribute only to $T(\bm{x}_{\rm
  objects}, \bm{x}_{\rm true})$ and neither depend on the new physics parameters $(A,
\bm{\alpha})$ nor change the interpretation of $T(\bm{x}_{\rm
  objects}, \bm{x}_{\rm true})$ as a normalized probability distribution.  
We note also that procedures like CKKW~\cite{Catani:2001cc,
  Krauss:2002up} or MLM~\cite{Mangano:2006rw}
matching, which attempt to combine aspects of the parton shower  with
event generation from matrix elements do not change the overall thrust
of our argument.  At most, we must modify our reweighting procedure to
use events $\bm{x}_{i,{\rm matched}}$ and generalized transfer
function $T(\bm{x}_{i,{\rm objects}}, \bm{x}_{i,{\rm matched}})$ in
Eq.~(\ref{eq:T}); these modifications, and the associated modifications to the
reweighting procedure for these events, are relatively straightforward.  

\section{Reweighting}
\label{sec:reweighting}

Given a set of unweighted hard-process events,
$\bm{x}_{i,~\rm{true}}$, generated with, e.g.,
{\tt MadGraph}, and following the facts about the generalized transfer
function $T$ from Eq.~(\ref{eq:T}) listed in
Section~\ref{sec:intro} and/or the discussion in
Section~\ref{sec:mc}, we can produce a set of unweighted
events $\bm{x}_{i,\rm{objects}}$ through MC simulation that
correspond to the hard-process events $\bm{x}_{i,\rm{true}}$.

These events can be used to obtain detector-level cross sections and
distributions, in a manner exactly analogous to that specified above
(c.f. Eq.~(\ref{eq:rho-mc})); namely if $N_{\rm
  detector}$ of the $N_{\rm generator}$ events which we simulate
satisfy our cuts on our
detector-simulated reconstructed physics objects, then our cross
section (times efficiencies and acceptances) is given by

\begin{equation}\label{eq:sigma-mc-detector}
\sigma_D = \sigma_G \frac{N_{\rm detector}}{N_{\rm generator}}.
\end{equation}
The content of the bin in the histogram for an arbitrary kinematic
variable $V$ containing values in the range $[V_{\rm min}, V_{\rm
  max}]$ is given by

\begin{equation}\label{eq:rho-mc-detector}
  \int_{V_{\rm min}}^{V_{\rm max}} \rho(V)\, dV \approx \frac{1}{N} \sum_i \left\{
  \theta(V(\bm{x}_{i, \rm objects}) - V_{\rm min}) - \theta( V(\bm{x}_{i, \rm objects}) - V_{\rm max})\right\}.
\end{equation}

Of course, these detector-level events correspond to the
generator-level events $\{\bm{x}_{i,\rm true}\}$, which were generated
using the new physics theory point $(A, \bm{\alpha})$.   
This raises the question: what if we now wish to obtain detector-level cross section
and observables for a different new physics model theory point, which we label $(B, \bm{\beta})$?  
Obviously we could simply perform the procedure described above on a new set of
unweighted, parton-level events generated for $(B,\bm{\beta})$.  This
approach, however, can become impractical very quickly, 
as fullsim detector simulation is very slow (rates on the order of minutes per
event are typical) relative to the rest of the simulation process.  

Instead, we note that the Eqs.~(\ref{eq:sigma-mc-detector}) and
(\ref{eq:rho-mc-detector}) can be thought of as 
the MC evaluation of cross sections or histograms using the
differential cross section for $\bm{x}_{\rm objects}$, i.e.

\begin{equation}\label{eq:P-A}
\frac{d\sigma(A,\bm{\alpha}; \bm{x}_{\rm objects})}{d \bm{x}_{\rm
    objects}} = \int T(\bm{x}_{\rm objects},
\bm{x}_{\rm true}) \frac{d\sigma(A,\bm{\alpha}; \bm{x}_{\rm true})}{d \bm{x}_{\rm
    true}} d\bm{x}_{\rm true}.
\end{equation}
If we replace the theory point $(A,\bm{\alpha})$
with $(B,\bm{\beta})$, we have instead

\begin{eqnarray}\label{eq:P-B}
\frac{d\sigma(B,\bm{\beta}; \bm{x}_{\rm objects})}{d \bm{x}_{\rm
    objects}}  & = & 
 \int T(\bm{x}_{\rm objects},
\bm{x}_{\rm true}) \frac{d\sigma(B,\bm{\beta}; \bm{x}_{\rm true})}{d \bm{x}_{\rm
    true}} d\bm{x}_{\rm true}
    \\ \nonumber
&=& \int T(\bm{x}_{\rm objects},
\bm{x}_{\rm true}) \frac{d\sigma(A,\bm{\alpha}; \bm{x}_{\rm true})}{d \bm{x}_{\rm
    true}} 
\,    R(A,\bm{\alpha},B,\bm{\beta}) \,
d\bm{x}_{\rm true} ,
\end{eqnarray}
with

\begin{equation}
R(A,\bm{\alpha},B,\bm{\beta}) \equiv  \bigg(
\frac{d\sigma(B,\bm{\beta}; \bm{x}_{\rm true})}{d \bm{x}_{\rm
    true}} \bigg/
\frac{d\sigma(A,\bm{\alpha}; \bm{x}_{\rm true})}{d \bm{x}_{\rm
    true}}
\bigg).
\label{eq:R}
\end{equation}
When we, e.g., perform the integral in Eq.~(\ref{eq:rho-mc-detector}) by
  generating unweighted events for the hard-process and then simulating,
  we are really replacing some region of $\bm{x}_{\rm true}$ space, with volume $V_i$, 
  with the event $\bm{x}_{i, \rm true}$.  The weight of this region
  in the calculation of quantities, is, approximately 
  $V_i \frac{d\sigma(A,\bm{\alpha}; \bm{x}_{\rm true})}{d \bm{x}_{\rm true}}$.  
  (These weights are equal for unweighted events, up to
  statistical fluctuations.  $V_i$ is small when the differential cross
  section is large; equivalently the number of events selected in a
  region of phase space is proportional, in the limit of large
  statistics, to the differential cross section in that region of phase
  space.)  When we then perform an integral like that in
  Eq.~(\ref{eq:rho-mc-detector}) using
  MC methods, we therefore must replace the weights:

 \begin{eqnarray}
(V_i) \frac{d\sigma(A,\bm{\alpha}; \bm{x}_{\rm true})}{d \bm{x}_{\rm
    true}}
\to
(V_i) \frac{d\sigma(B,\bm{\beta}; \bm{x}_{\rm true})}{d \bm{x}_{\rm
    true}}
\equiv
(V_i) \frac{d\sigma(A,\bm{\alpha}; \bm{x}_{\rm true})}{d \bm{x}_{\rm
    true}}
R(A,\bm{\alpha},B,\bm{\beta}).
\end{eqnarray}

At the level of MC events, this replacement corresponds to weighting
each event by the ratio of hard-process differential cross sections (\ref{eq:R}).
With this procedure we obtain, for cross sections

\begin{equation}\label{eq:sigma-mc-detector-weighted}
\sigma_D(B,\bm{\beta}) =\sigma_G(A,\bm{\alpha}) \,  \frac{1}{N_{\rm generator}} \sum_{i, \rm (accepted)} R(A,\bm{\alpha},B,\bm{\beta}).
\end{equation}
and for histograms

\begin{eqnarray}\label{eq:rho-mc-detector-weighted}
  \int\limits_{V_{\rm min}}^{V_{\rm max}} \!\!\!\! \rho(V) dV \approx  \!\!\!\!\!\!
  \sum\limits_{i, \rm accepted} \!\!\!\! \frac{R(A,\bm{\alpha},B,\bm{\beta})}{N_{\rm generator}}
  \biggl[\theta(V(\bm{x}_{i, \rm objects}) - V_{\rm min}) - \theta(V(\bm{x}_{i, \rm objects})- V_{\rm max} )\biggr].  
\end{eqnarray}

\section{Applications}
\label{sec:applications}

%%%%%%%%%%%%% Beginning OF FIGURE ################%%%%%%%%%%%%%
\begin{figure}[t]
\begin{center}
\includegraphics[width=0.8\textwidth]{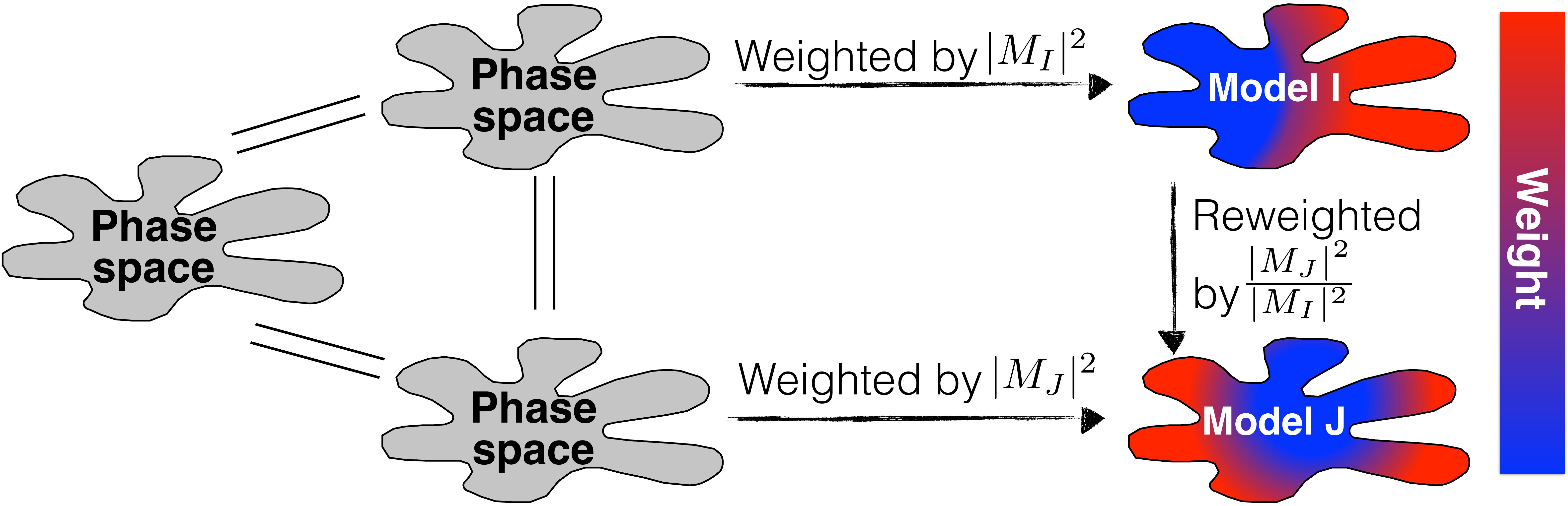} 
\end{center}
\vspace{-20 pt}
\caption{A conceptual diagram describing the reweighting procedure. 
  Assume that we have generated MC events for a particular event topology in physics
  model, $I$.  If another model, $J$, also allows this
  topology (with equal masses for all final state particles and narrow resonances)
  we can also perform MC analyses in the context of model $J$ without generating
  new events. The idea is to reweight the existing MC event samples, generated for 
  model $I$, by a factor of $\frac{|M_J|^2}{|M_I|^2}$, where $|M_I|^2$ ($|M_J|^2$)
  is the squared matrix element for the given parton-level event in
  model $I$ (model $J$).
  \label{fig:reweight}}
\end{figure}
%%%%%%%%%%%%% End OF FIGURE %%%%%%%%%%%%%%%%%%%%%%%%%%%%%%%%%
 
In this section, we demonstrate the use of reweighting methods through
three example analyses. We also discuss the bin-by-bin errors in histograms obtained
by reweighting. In Subsection~\ref{sec:h4l} we will show how 
to reweight events to study different coupling structures for Higgs to four-lepton events.  
This will be followed, in Subsection~\ref{sec:antler}, by an investigation of
reweighting to obtain distributions for a UED model from MC samples generated
for a SUSY model.  Finally, in Subsection~\ref{sec:clustering} we will study the effects of
showering, hadronization, (fast) detector simulation, and jet reconstruction.

Especially for the benefit of readers who may have skipped
Section~\ref{sec:mc}, in Fig.~\ref{fig:reweight} we present a cartoon which describes the
reweighting procedure. 
Essentially we can obtain distributions in some physics model $J$ by  
reweighting MC events generated for a model $I$.  The reweighting
factor (\ref{eq:R}) is given by the ratio of differential cross sections evaluated
for the ``truth''-level parton-level MC event in each model.  In the
limit where model $I$ and model $J$ are produced from the same initial
partons and have the same masses for final state particles, the parton
distribution functions and
phase space factors in this ratio cancel, and we are left with the
ratio of squared matrix elements, as described in the caption to
Fig.~\ref{fig:reweight}.
The different colorations of the cartoons representing model $I$ and model $J$ 
in Fig.~\ref{fig:reweight} represent the different distributions of the 
squared matrix elements over the common phase space in the two models.

In the examples provided, the factors used for
  reweighting were obtained using the ``standalone'' matrix element
  calculating code which can be generated automatically, in either
  {\tt Fortran} or {\tt C++}, from {\tt MadGraph5}~\cite{Alwall:2011uj}.  
  This code requires the external parton momenta, which should be
  provided by a short, user-supplied, ``wrapper'' code. 
  One can also generate similar standalone code in {\tt C} or {\tt Mathematica}
  using {\tt CalcHEP} 
  or {\tt CompHEP}~\cite{Belyaev:2012qa, Pukhov:1999gg, Boos:2004kh}.  
  We encourage the authors of these and other similar automatic matrix element
  calculators to further increase the user-friendliness of their tools for the purpose of reweighting.

Before proceeding, we pause to present a caveat about the
sorts of models one can study using reweighting.  Obviously we can only
use a reweighting procedure if the event being reweighted is possible
in both models.  So the final state particles must have the same
masses in both models.  Additionally, when one or both models contain
intermediate resonances, it is important that those resonances, if sufficiently
narrow, have the same masses in each model.  Otherwise, the
resulting extreme differences in the density of events in the phase
space of the two models can lead to undersampling of important regions
of phase space. However, this is not, in principle, an insurmountable difficulty,
and several practical approaches for dealing with undersampling will be 
discussed in Section~\ref{sec:practical}.

\subsection{Changing the Coupling Structure: Higgs to Four Leptons}
\label{sec:h4l}

%%%%%%%%%%%%% Beginning OF FIGURE ################%%%%%%%%%%%%%
\begin{figure}[t!]
\begin{center}
\includegraphics[width=0.85\textwidth]{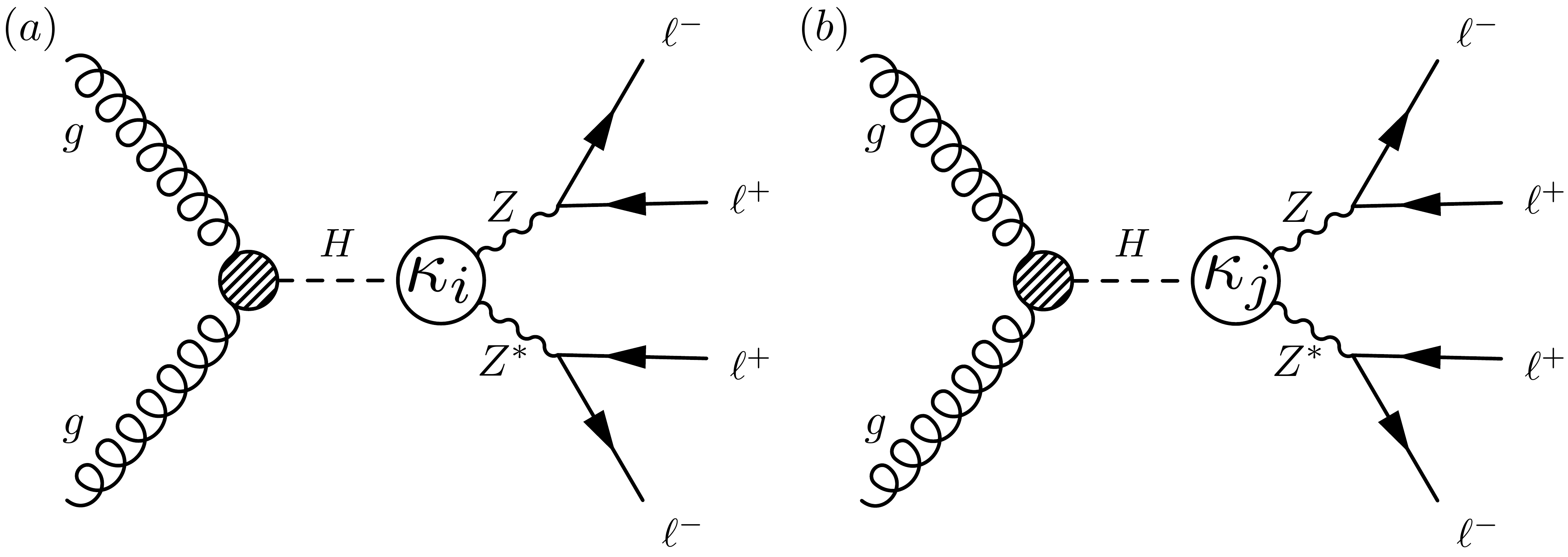} 
\end{center}
\vspace{-15 pt}
\caption{
  \label{fig:golden}
  Feynman diagrams showing the gluon fusion production of a Higgs
  boson, $H$, followed by its decay to two $Z$ bosons (at least one of
  which is off-shell) and the subsequent decay of the $Z$ bosons to leptons.
  Following Eq.~(\ref{Lag}), the coupling of the Higgs boson to a pair of $Z$
  bosons is labeled by $\kappa_i$ or $\kappa_j$.  }
\end{figure}
%%%%%%%%%%%%% End OF FIGURE %%%%%%%%%%%%%%%%%%%%%%%%%%%%%%%%%

The ``golden'' Higgs to four-lepton channel (see Fig.~\ref{fig:golden})
is quite sensitive,
both for the discovery and for the subsequent measurement
of the spin and parity properties of the Higgs
boson~\cite{Dell'Aquila:1985ve, Nelson:1986ki, Kniehl:1990yb,
 Soni:1993jc, 
Chang:1993jy, Barger:1993wt, Arens:1994wd, Choi:2002jk, 
Allanach:2002gn, Buszello:2002uu, Godbole:2007cn, 
Kovalchuk:2008zz, Keung:2008ve, Antipin:2008hj, Cao:2009ah, Gao:2010qx, 
DeRujula:2010ys, Englert:2010ud, Matsuzaki:2011ch, DeSanctis:2011yc, 
logan, Gainer:2011xz, oai:arXiv.org:1110.4405, Englert:2012ct, 
Campbell:2012cz, Campbell:2012ct, Kauer:2012hd, Kniehl:2012rz, Moffat:2012pb, 
Coleppa:2012eh, Bolognesi:2012mm, Boughezal:2012tz, Stolarski:2012ps, 
Cea:2012ud, Kumar:2012ba, Geng:2012hy, Avery:2012um, Masso:2012eq, 
Chen:2012jy, Modak:2013sb, Kanemura:2013mc, Gainer:2013rxa, 
Isidori:2013cla, Frank:2013gca, Grinstein:2013vsa, Caola:2013yja, 
Banerjee:2013apa, Sun:2013yra, Anderson:2013afp, 
Chen:2013waa, Buchalla:2013mpa, Chen:2013ejz, Campbell:2013una, Chen:2014pia, 
Gonzalez-Alonso:2014rla, Gainer:2014hha, Chen:2014gka}, and therefore
plays an important role in the experimental study of the
Higgs at the LHC~\cite{Chatrchyan:2012jja, Aad:2013xqa, Chatrchyan:2013mxa,
  1523699, 1542341, 1542387}.
Assuming that the putative Higgs is spin-zero, we can
then characterize its couplings to $Z$ bosons,
following~\cite{Gainer:2014hha},
using the Lagrangian
\begin{eqnarray}
\mathcal{L} \supset
\sum_{i=1}^5 \kappa_i \mathcal{O}_i = &&
-\kappa_1 \frac{M_Z^2}{v} H Z_\mu Z^\mu
-\frac{\kappa_2}{2 v} H F_{\mu\nu} F^{\mu\nu}
-\frac{\kappa_3}{2 v} H F_{\mu\nu} \tilde{F}^{\mu\nu}  \nonumber \\
&&+\frac{\kappa_4 M_Z^2}{M_X^2 v} \Box H Z_\mu Z^\mu
+\frac{2\kappa_5}{v} H Z_\mu \square Z^\mu.
\label{Lag}
\end{eqnarray}
Therefore, in general, the $HZZ$ couplings involve five parameters (up to lowest non-trivial dimension).  
Following the ``Geolocating''
approach~\cite{Gainer:2013rxa}, we can remove the degree of freedom
associated with the partial width, leaving us with a ``sphere'' of
four-dimensions.  To explore this space fully, we may need to employ
reweighting procedures, such as those described here.

%%%%%%%%%%%%% Beginning OF FIGURE ################%%%%%%%%%%%%%
\begin{figure}[t]
\begin{center}
\includegraphics[width=0.328\textwidth]{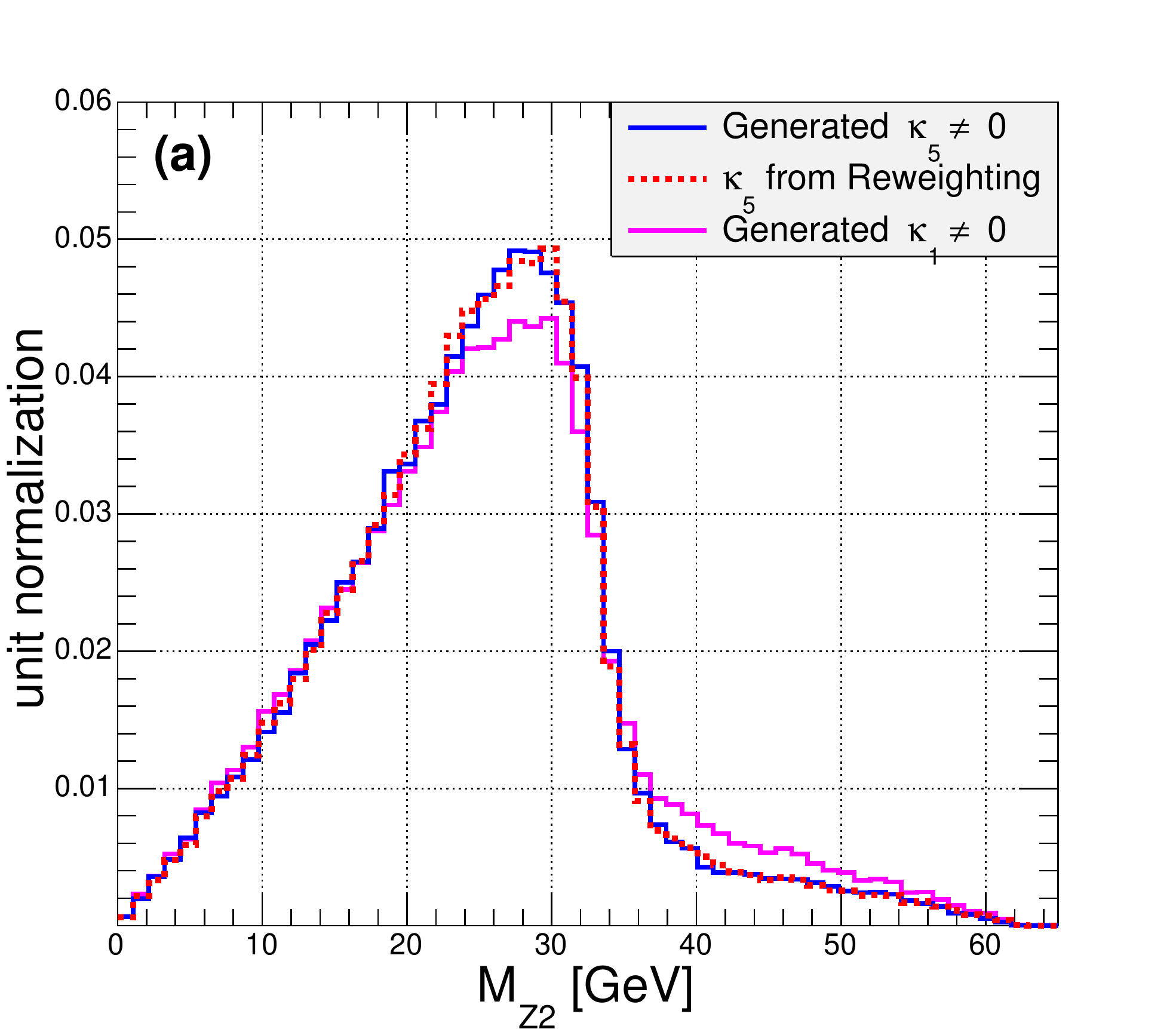} 
\includegraphics[width=0.328\textwidth]{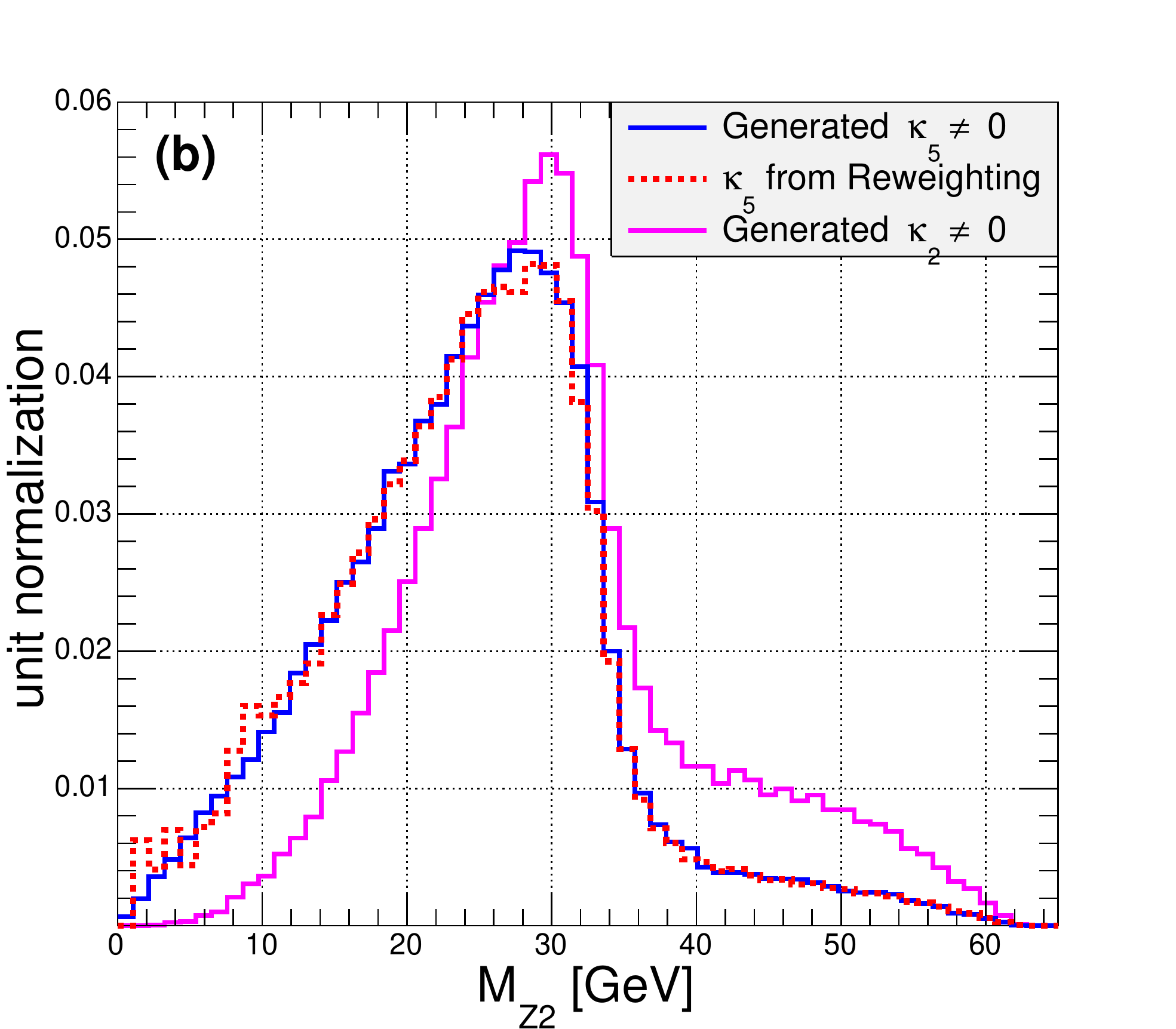}
\includegraphics[width=0.328\textwidth]{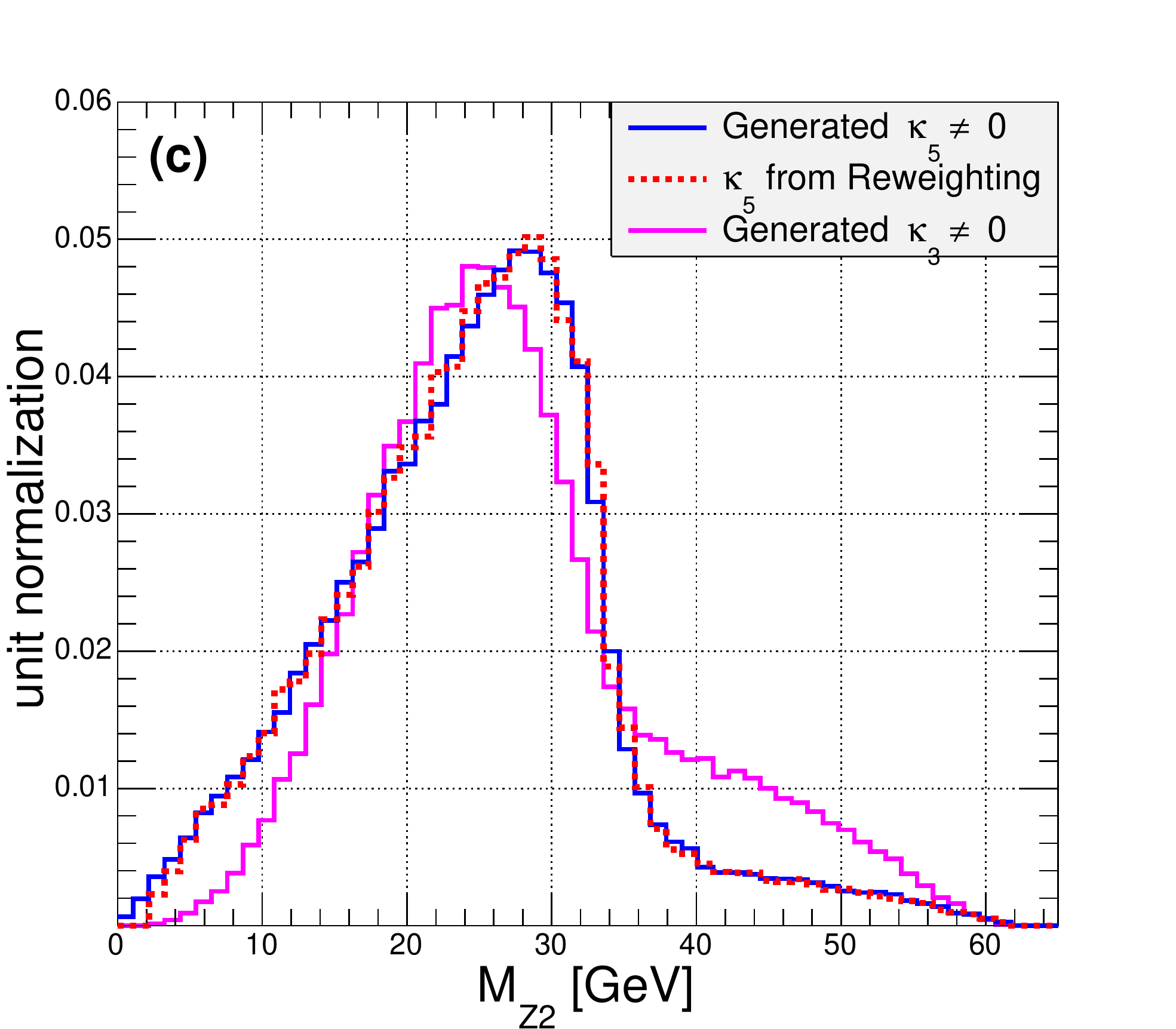}
\end{center}
\vspace{-20 pt}
\caption{
  \label{fig:MZ2}
  Normalized distributions of the quantity $M_{Z_2} =
  \min{(M_{e^+e^-},M_{\mu^+\mu^-})}$ at parton-level.  In each panel,
  the solid magenta distributions
  are obtained using events generated under various coupling
  hypotheses: a) $\kappa_i = (1,0,0,0,0)$, b)
  $\kappa_i = (0,1,0,0,0)$, and c) $\kappa_i = (0,0,1,0,0)$, while the solid blue 
   distribution is obtained by directly generating events for a model
   with $\kappa_i = (0,0,0,0,1)$.
  Alternatively, we can obtain the same $\kappa_5\ne 0$ distribution,
  by recycling the events used for the 
  solid magenta histograms, and reweighting them with the
  corresponding ratio of matrix elements squared.
  The results from such reweighting are shown by the dotted red distributions.}
\end{figure}
%%%%%%%%%%%%% End OF FIGURE %%%%%%%%%%%%%%%%%%%%%%%%%%%%%%%%%

As an example of such analyses, in Fig.~\ref{fig:MZ2} we study the distribution of the 
smaller dilepton invariant mass\footnote{$M_{Z_2}$ has been shown 
to be a relatively sensitive variable for signal versus background discrimination 
and for measuring Higgs properties, see, for example~\cite{Avery:2012um,Chen:2013waa}.} 
$M_{Z_2} = \min{(M_{e^+e^-},M_{\mu^+\mu^-})}\,$ in the $2e2\mu$
final state. In each of the three panels in Fig.~\ref{fig:MZ2}, the solid blue histogram is always the 
normalized $M_{Z_2}$ distribution for the pure $\kappa_5$ case, i.e.
when $\kappa_5\ne 0$ and all other $\kappa_i$ ($i=1,\ldots, 4$) are zero. 
On the other hand, the solid magenta histograms correspond to the case when only 
one of the other $\kappa_i$ couplings is turned on:  $\kappa_1\ne 0$ in panel (a),
$\kappa_2\ne 0$ in panel (b), and $\kappa_3\ne 0$ in panel (c).
Our reweighting procedure now allows us to obtain the shape of the 
solid blue histogram by reweighting the corresponding magenta plot.
The results are shown by the dotted red lines in Fig.~\ref{fig:MZ2}.
By comparing the solid blue and dotted red distributions, we
observe a very good match, which validates our procedure.

  However, it is not sufficient to show that the central bin values
  for each histogram are reproduced by the reweighting procedure.
  To perform statistical analyses using the reweighted histograms, we
  need to understand the error (or uncertainty) on the histogram bin values.  
  Following Ref.~\cite{Lyons:1986em}, if
  there are $N$ events in a bin, each of which has weight $w_k$, then
  the value of that bin in the reweighted histogram is
\begin{equation}
\label{eq:T2}
T = \sum w_k \equiv N \langle w \rangle.
\end{equation}
The error on $T$ is given by
\begin{equation}
\label{eq:delta}
\delta = \sqrt{\sum w_k^2} = \sqrt{N} \sqrt{\langle w^2 \rangle}.
\end{equation}
Clearly, in the special case of an unweighted histogram, $\delta$ yields
$\sqrt{N}$ which reproduces the well-known expression from Poisson
statistics.  Combining Eqs.~(\ref{eq:T2}) and~(\ref{eq:delta}), and writing
$\langle w^2 \rangle$ in terms of the variance on the weights,
\begin{equation}
\sigma_w^2 = \langle w^2 \rangle - \langle w \rangle^2,
\label{eq:variance}
\end{equation}
we find the fractional error on the bin value in a weighted
histogram to be given by
\begin{equation}
\label{eq:weighted-error}
\frac{\delta}{T} = \frac{1}{\sqrt{N}} \sqrt{1 +
  \frac{\sigma_w^2}{\langle w \rangle^2}}.
\end{equation}
For the case of interest to us, the $w_k$ are the reweighting factors
$R(A,\bm{\alpha},B,\bm{\beta})$ used to reweight events generated for
model $A$, in order to obtain a histogram for model $B$. (Note that 
the value of $N$ is the same for the two models.)  
Since in general the reweighting factor varies from event to event, the variance (\ref{eq:variance})
is nonzero. Thus, when we reweight, the statistical error increases, 
and Eq.~(\ref{eq:weighted-error}) quantifies this effect.

%%%%%%%%%%%%% Beginning OF FIGURE ################%%%%%%%%%%%%%
\begin{figure}[th]
\begin{center}
\includegraphics[width=0.328\textwidth]{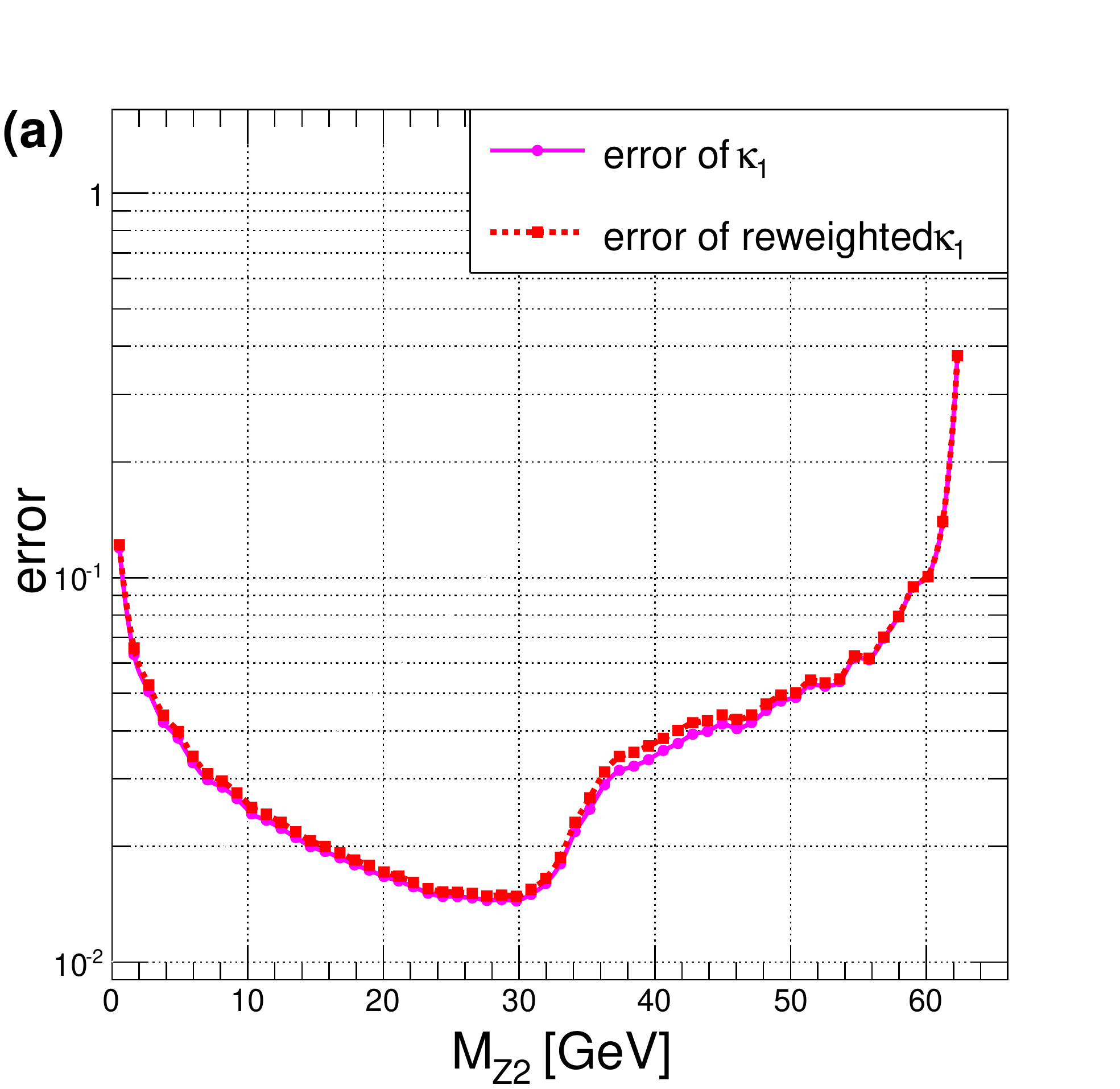} 
\includegraphics[width=0.328\textwidth]{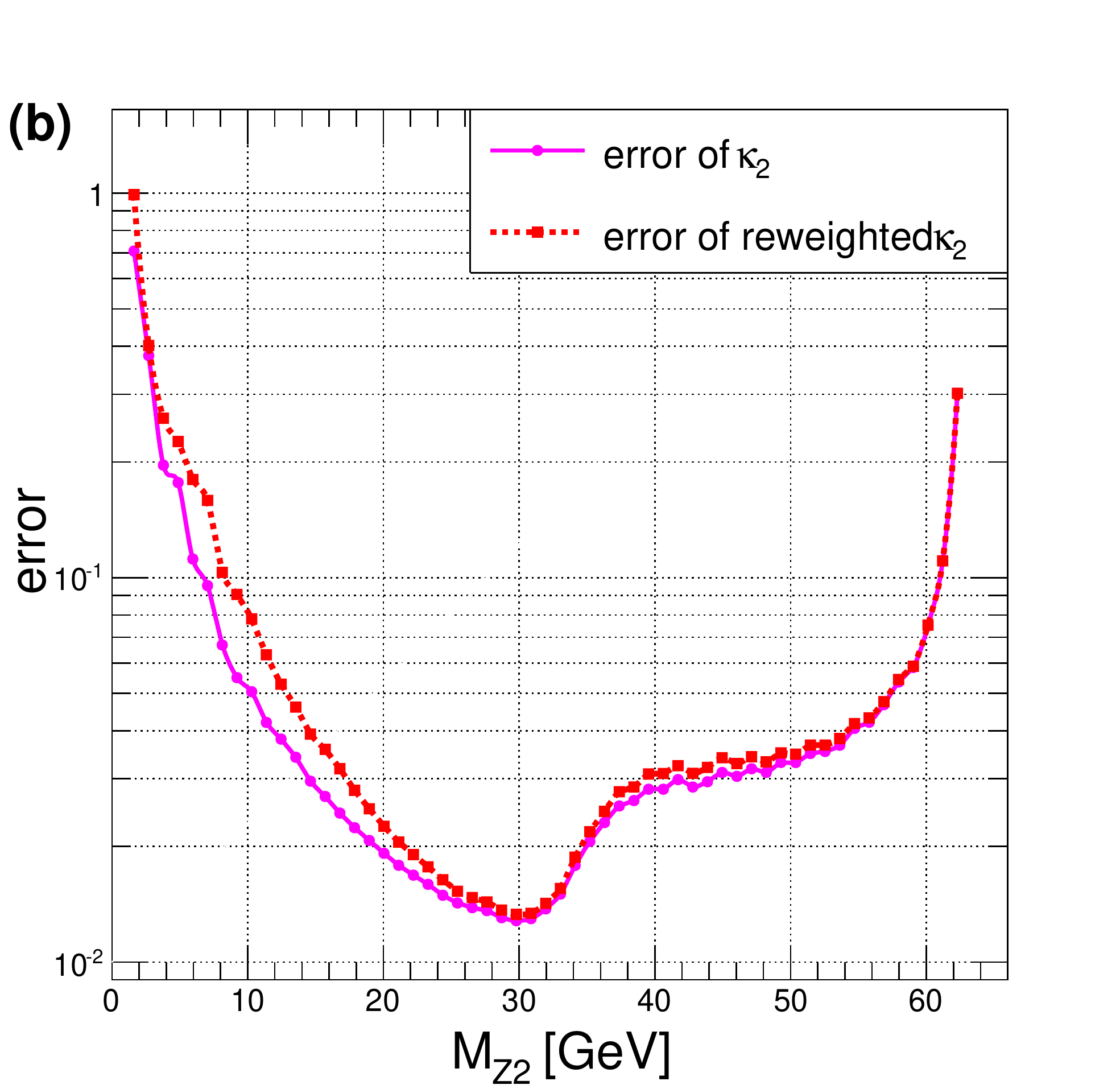}
\includegraphics[width=0.328\textwidth]{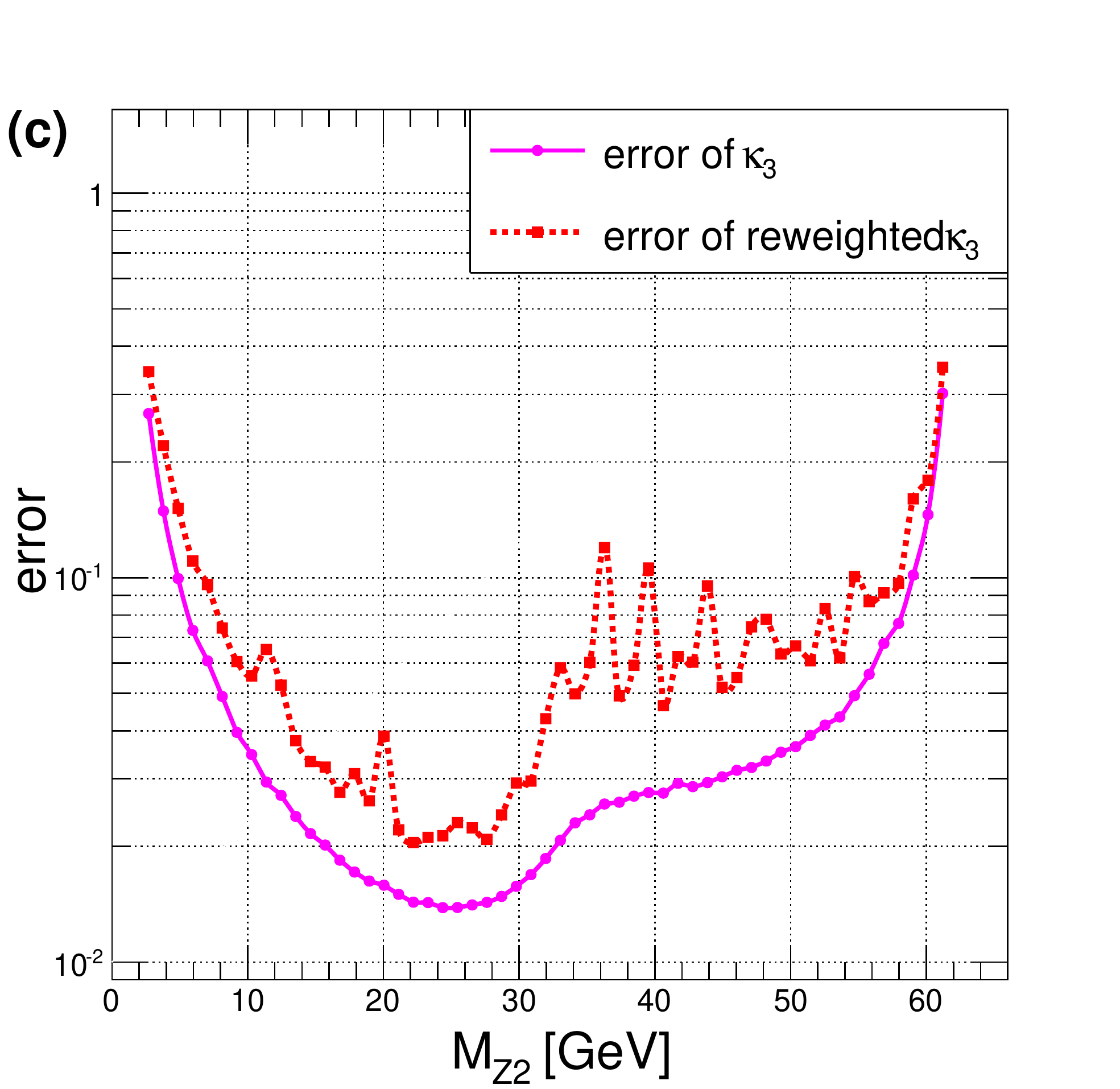}
\end{center}
\vspace{-20 pt}
\caption{
  \label{fig:error}
The fractional error on each $M_{Z2}$ bin value, obtained by applying Eq.~(\ref{eq:weighted-error}) to the
red-dashed and magenta histograms from Fig.~\ref{fig:MZ2}.
}
\end{figure}
%%%%%%%%%%%%% End OF FIGURE %%%%%%%%%%%%%%%%%%%%%%%%%%%%%%%%%

\begin{figure}[th]
\begin{center}
\includegraphics[width=0.328\textwidth]{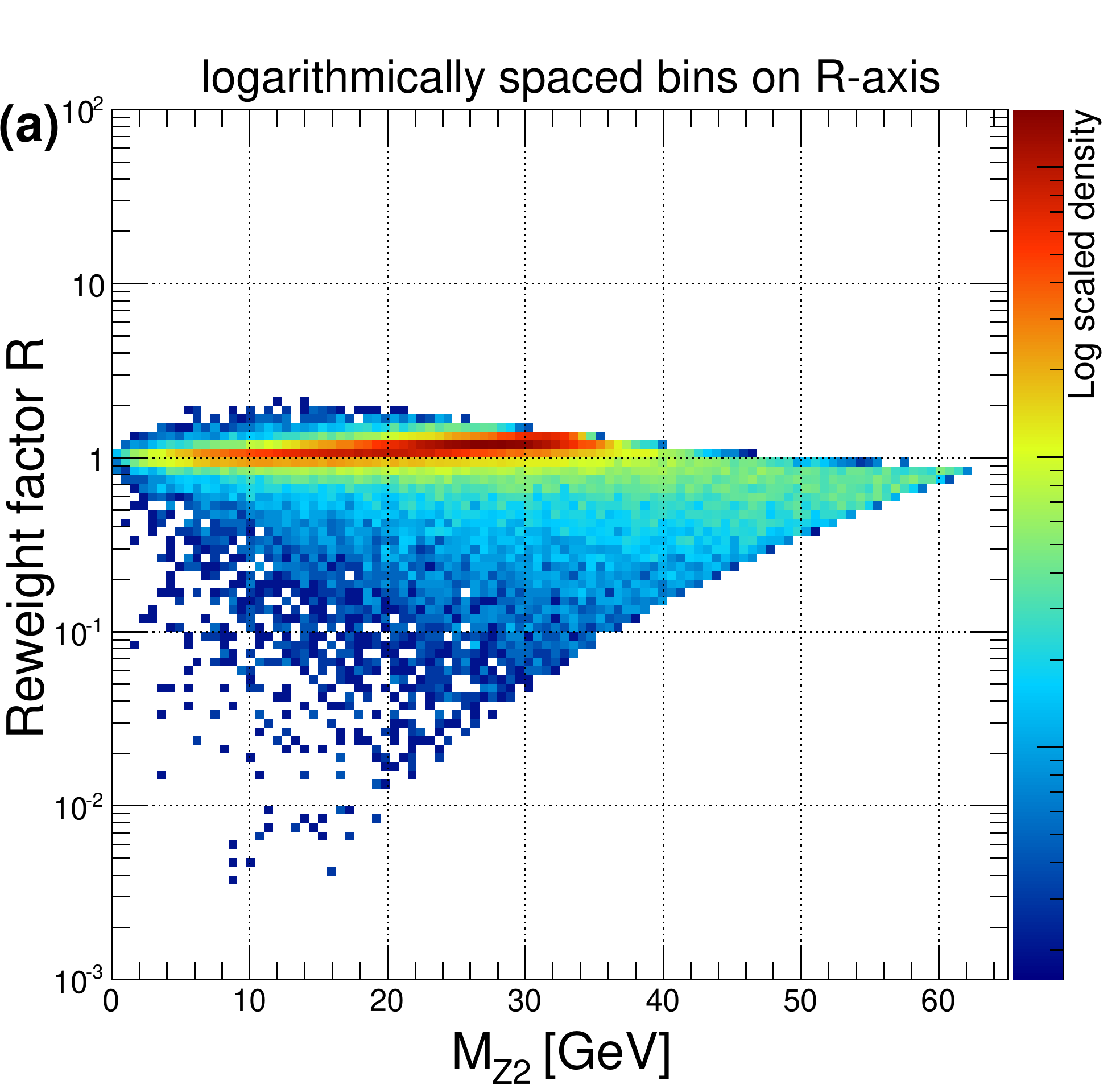} 
\includegraphics[width=0.328\textwidth]{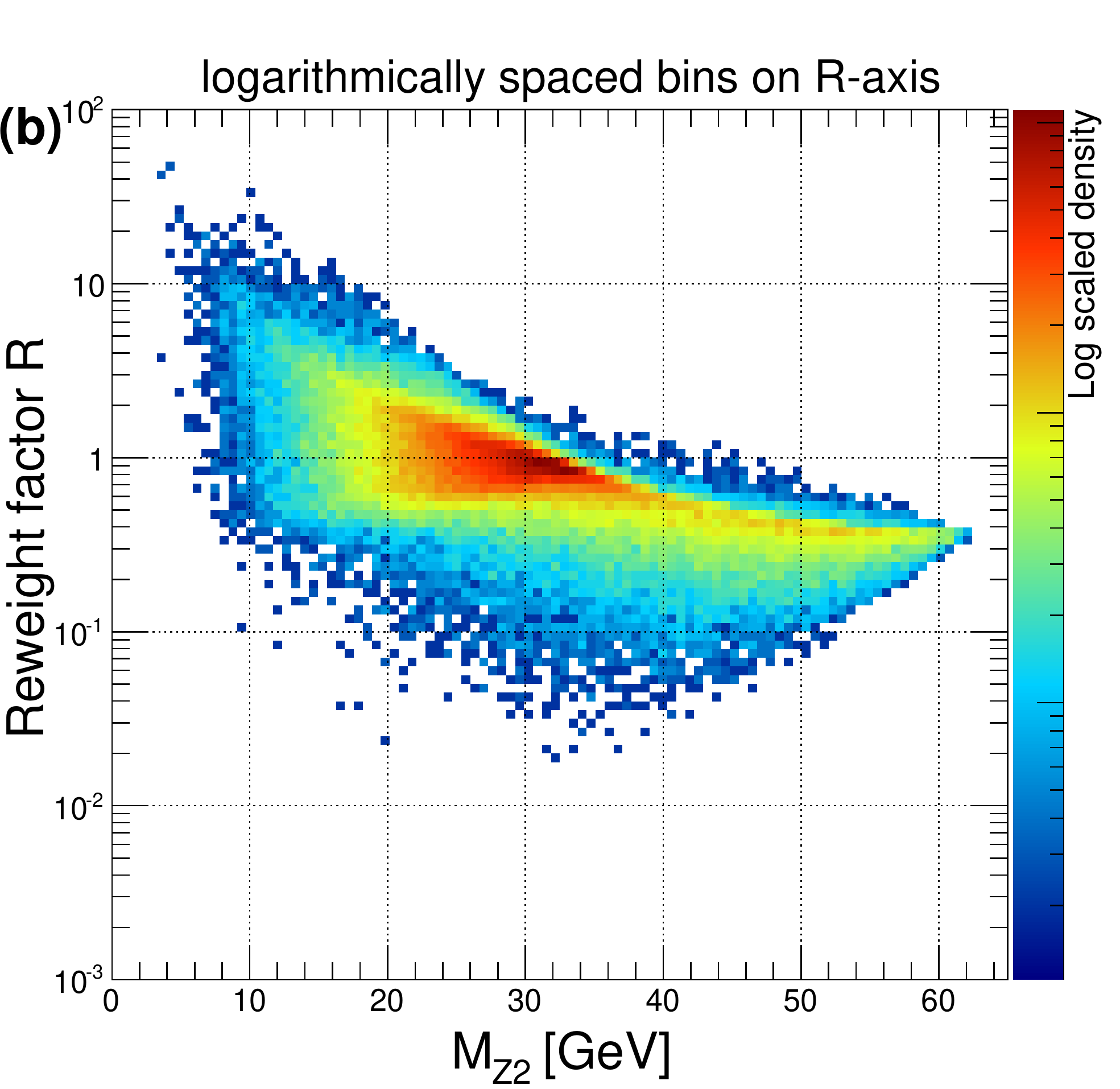}
\includegraphics[width=0.328\textwidth]{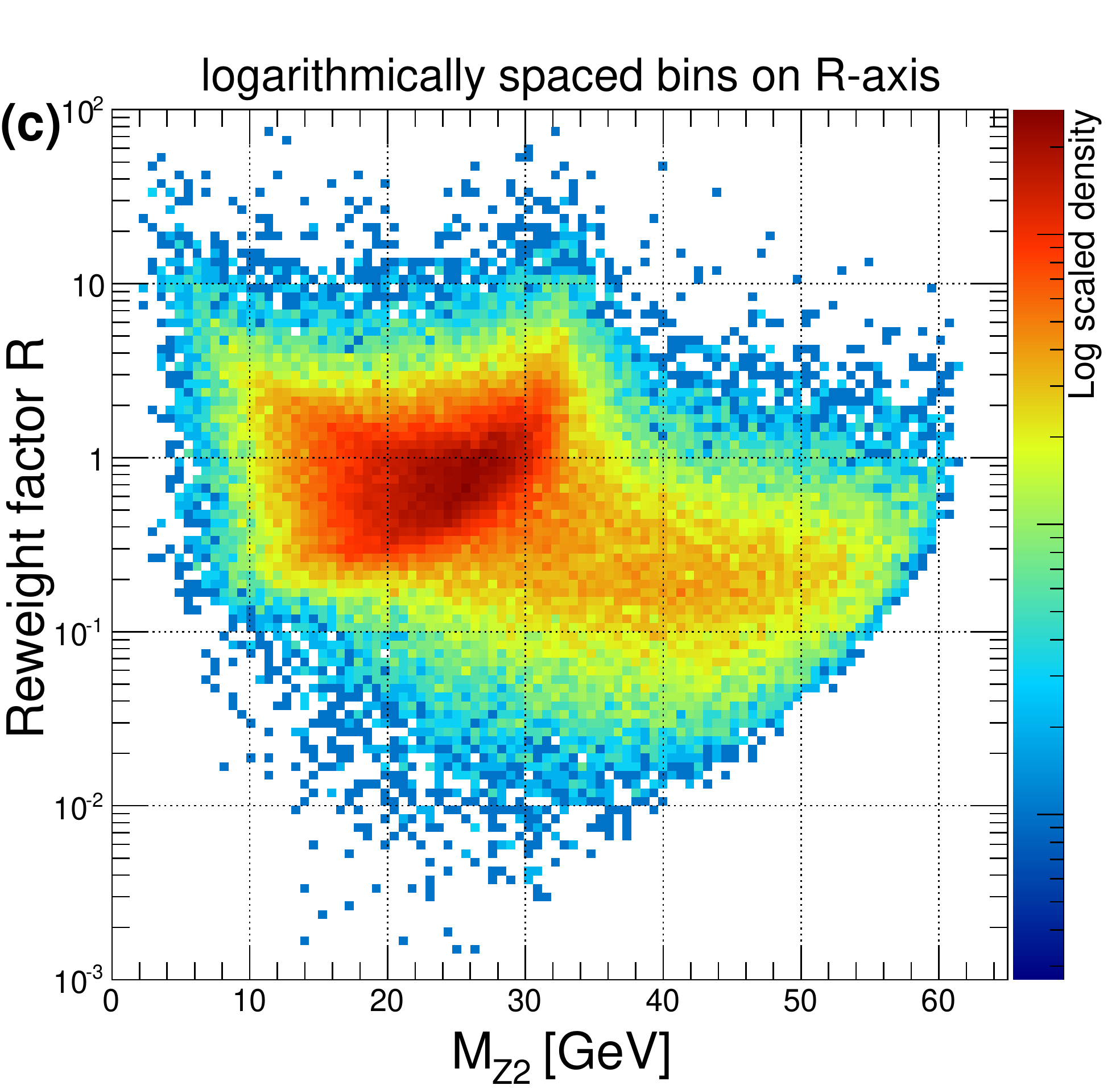}
\end{center}
\vspace{-20 pt}
\caption{
  \label{fig:R}
The values of the reweighting factor, $R(A,\bm{\alpha},B,\bm{\beta})$,
and $M_{Z_2}$ for the events used to obtain the histograms in Fig.~\ref{fig:MZ2}.
}
\end{figure}

To provide a concrete demonstration of these effects, in Fig.~\ref{fig:error}(a-c) 
we show the fractional error, as calculated using Eq.~(\ref{eq:weighted-error}), on
each bin of the corresponding histograms from Fig.~\ref{fig:MZ2}(a-c).
In Fig.~\ref{fig:error}, solid magenta lines denote the fractional error  (which scales as $1/\sqrt{N}$)
in the original model with unweighted events,
while the red dotted lines show the error on the corresponding histogram
obtained from the same events using reweighting. In accordance with
Eq.~(\ref{eq:weighted-error}), the error is in general greater for the 
reweighted histogram, but not always --- e.g., in Fig.~\ref{fig:error}(a), the
errors before and after reweighting are essentially the same.  
%Also, the extent to which this is the case varies considerably within a given histogram.  
To better illustrate the origin of the errors after reweighting, in Fig.~\ref{fig:R}
we provide two dimensional temperature plots of the reweighting factor, $R$, 
%as defined by Eq.~(\ref{eq:R}), 
and the corresponding value of $M_{Z2}$ for each event. 
We see a clear correlation between the spread in the values of $R$ 
and the magnitude of the increase in the fractional errors in Fig.~\ref{fig:error}.

%%%%%%%%%%%%% Beginning OF FIGURE ################%%%%%%%%%%%%%
%%%%%%%%%%%%% End OF FIGURE %%%%%%%%%%%%%%%%%%%%%%%%%%%%%%%%%

At this point, one might wonder how the errors on histograms which were reweighted
from an initial model $A$ to another model $B$ compare to the errors on histograms 
built from unweighted events generated directly in model $B$.
Since the error on the bin values in the unweighted histogram is $1/\sqrt{T}$, 
from Eqs.~(\ref{eq:T2}) and (\ref{eq:weighted-error}) we get
\begin{equation}
\label{eq:reweighting-error} 
\bigg(\frac{\delta}{T}\bigg) / \bigg(\frac{1}{\sqrt{T}}\bigg) =
\sqrt{\langle w \rangle} \sqrt{1 +
  \frac{\sigma_w^2}{\langle w \rangle^2}},
\end{equation}
where we assume sufficiently large statistics, so that the number of unweighted events 
is close to the expected value $T$. We see that the second factor in Eq.~(\ref{eq:reweighting-error}) 
always leads to an increase in the error when reweighting, in agreement with Fig.~\ref{fig:error}.
On the other hand, the first factor, $\sqrt{\langle w \rangle}$, can be either larger or smaller than 1,
depending on the relative weights between models $A$ and $B$.  Using
Eqs.~(\ref{eq:weighted-error}) and~(\ref{eq:reweighting-error}), one can
quantify the error from reweighting in the region of interest. If that error turns out to be 
unacceptably large for the task at hand, one can apply the procedures discussed below in Sec.~\ref{sec:practical} 
to further reduce those errors.

In Fig.~\ref{fig:MZ2} we presented distributions for the $M_{Z_2}$
variable, which we studied further in Figs.~\ref{fig:error} and~\ref{fig:R}.
However, we wish to emphasize that the
variable whose distribution is obtained via reweighting could also be
an optimized multivariate analysis-based variable such as the MELA
KD~\cite{Gao:2010qx,Chatrchyan:2012jja}, MEKD~\cite{Avery:2012um}, or
the output of a boosted
decision tree or neural network analysis.  Hence reweighting can be
used in tandem with sophisticated and powerful multivariate analysis
methods~\cite{Bhat:2010zz}.

\subsection{Changing Spin Assignments: The Antler Topology}
\label{sec:antler}

%%%%%%%%%%%%% Beginning OF FIGURE ################%%%%%%%%%%%%%
\begin{figure}[t!]
\begin{center}
\includegraphics[width=0.85\textwidth]{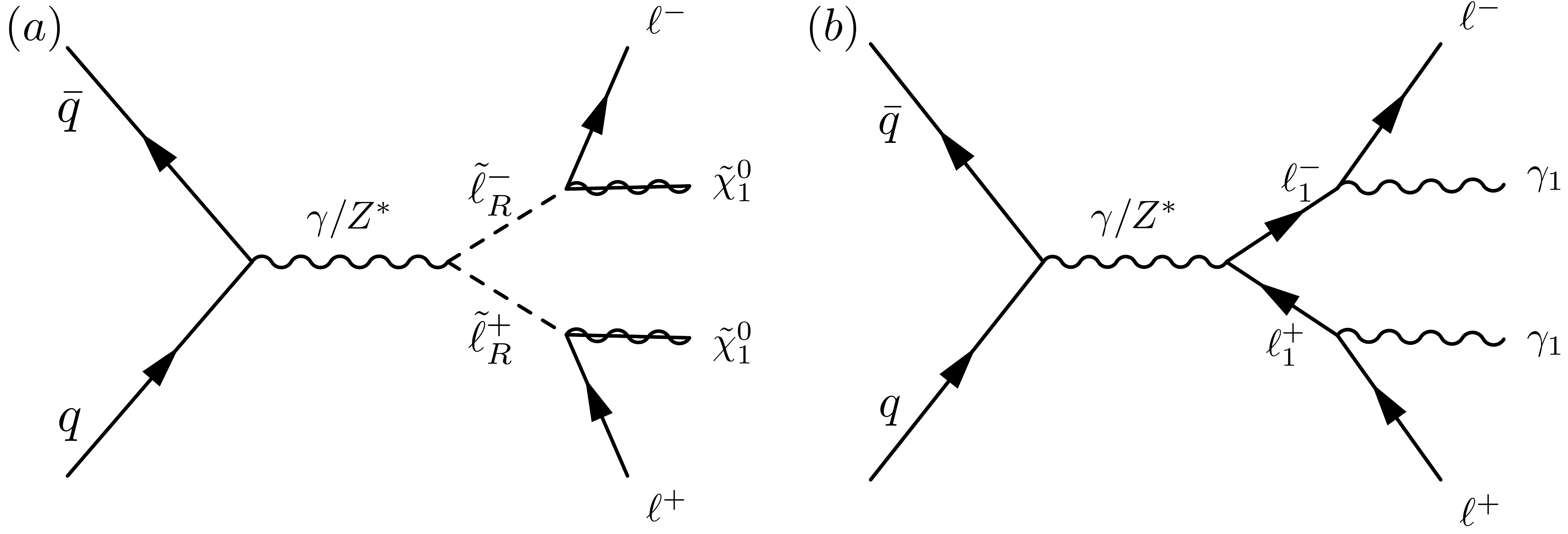} 
\end{center}
\vspace{-15 pt}
\caption{
  Feynman diagrams for the antler topology in a) SUSY and b) MUED.  
  We can study the MUED spin assignment 
  in this topology by reweighting SUSY events, as demonstrated in the text.\label{fig:antler}}
\end{figure}
%%%%%%%%%%%%% End OF FIGURE %%%%%%%%%%%%%%%%%%%%%%%%%%%%%%%%%

Extensive simplified model searches~\cite{ArkaniHamed:2007fw,Alwall:2008ag,Alves:2011wf}
have been performed by the ATLAS and CMS collaborations in various
channels~\cite{ATLAS:2011lka,Okawa:2011xg,ATLAS:2011ffa,Chatrchyan:2013sza}.
When performing a simplified model analysis, one generally 
fixes the spins of the particles involved in the simplified event topology.
One common choice is to use SUSY spin assignments, though another 
possibility is to decay all particles by pure phase space.
It is important to be able to interpret the results from these
searches in the context of other models, with different spin assignments.
A particularly well motivated example of an alternative spin assignment is provided by 
the minimal UED model (MUED) \cite{Cheng:2002iz}.

To perform reinterpretations for different spin assignments, we need to recalculate 
cross sections, branching ratios, and efficiencies for the given study point in the new model. 
We can obtain cross sections and branching ratios for the new model
point via theoretical formulae, but efficiencies generally must be
obtained through MC simulation.  We emphasize that we can recycle
existing MC samples
to determine these efficiencies for different new physics models in an
efficient manner.  In this subsection, we demonstrate the utilization
of MC event samples generated for a SUSY model to perform a collider
analysis of the MUED model, in the context of the
``antler'' topology  \cite{Han:2009ss,Edelhauser:2012xb}
depicted in Fig.~\ref{fig:antler}.  In particular we show 
distributions of collider observables, which could be used
to discriminate between SUSY and MUED~\cite{Battaglia:2005zf,Datta:2005zs,Meade:2006dw,Athanasiou:2006ef,Wang:2006hk,Burns:2008cp}.
Numbering the leptons as $1$ for the $\ell^-$ and $2$ for the $\ell^+$,
these variables are:

%%%%%%%%%%%%% Beginning OF FIGURE ################%%%%%%%%%%%%%
\begin{figure}[t!]
\begin{center}
\includegraphics[width=0.328 \textwidth]{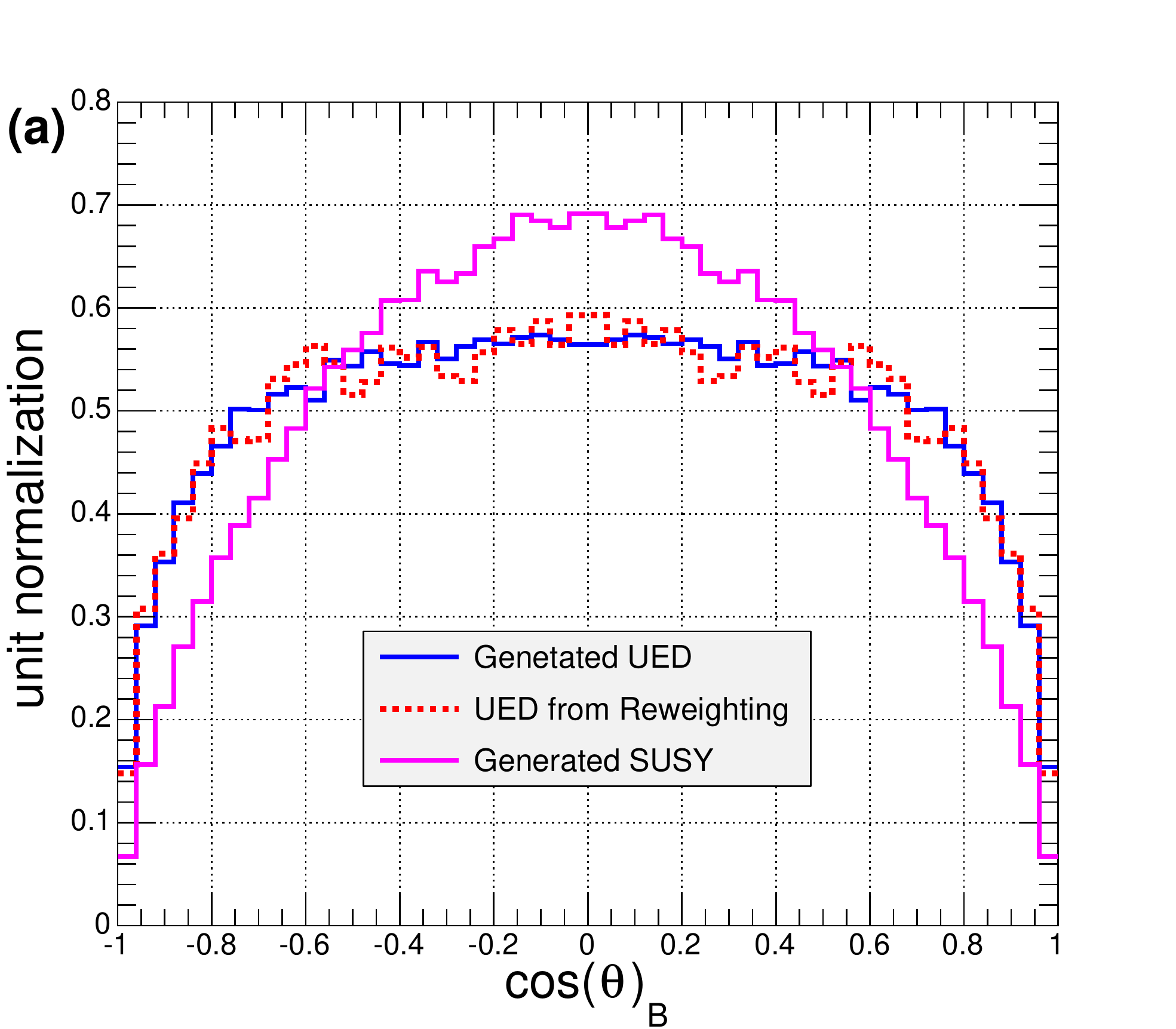} 
\includegraphics[width=0.328 \textwidth]{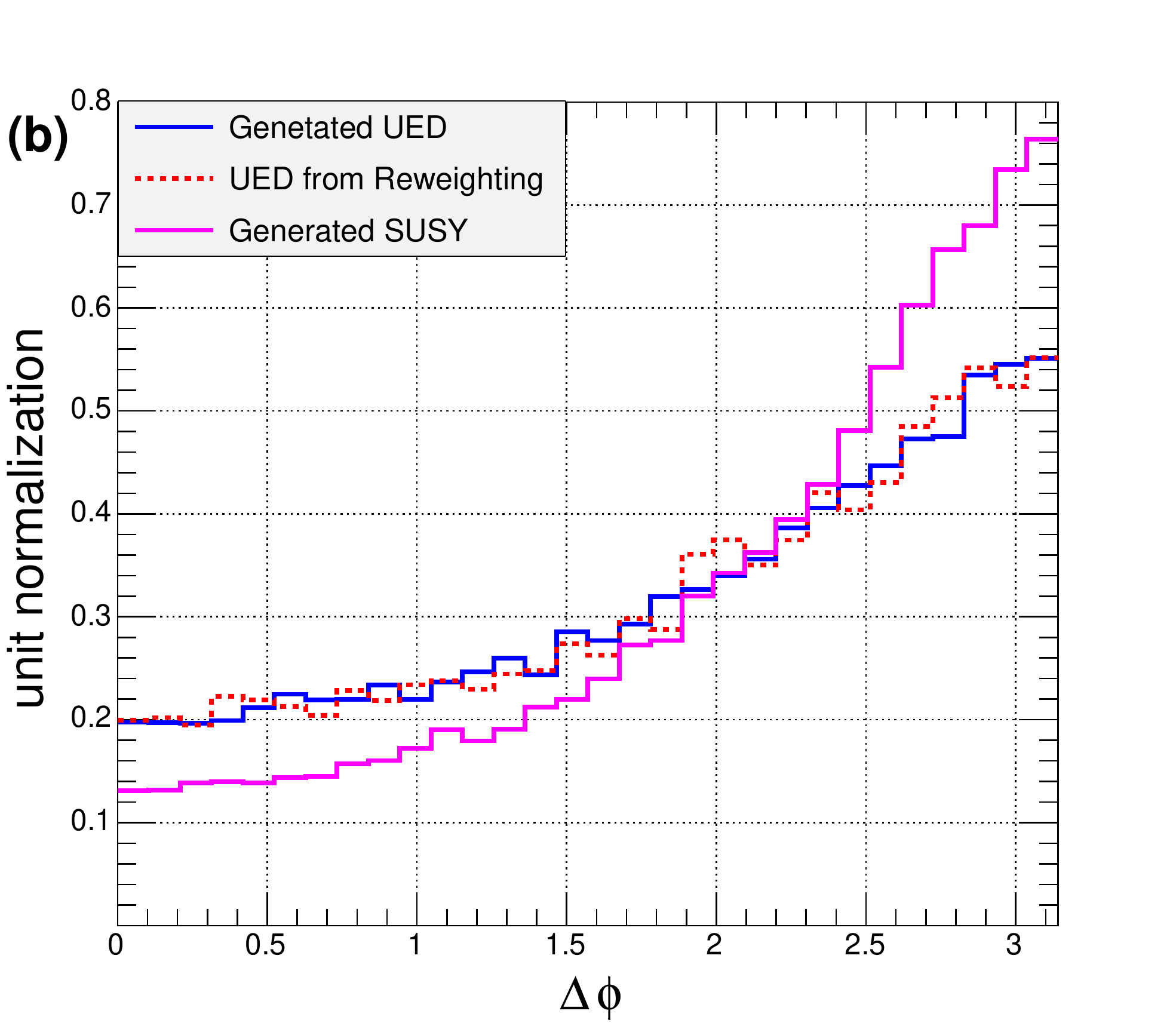}
\includegraphics[width=0.328 \textwidth]{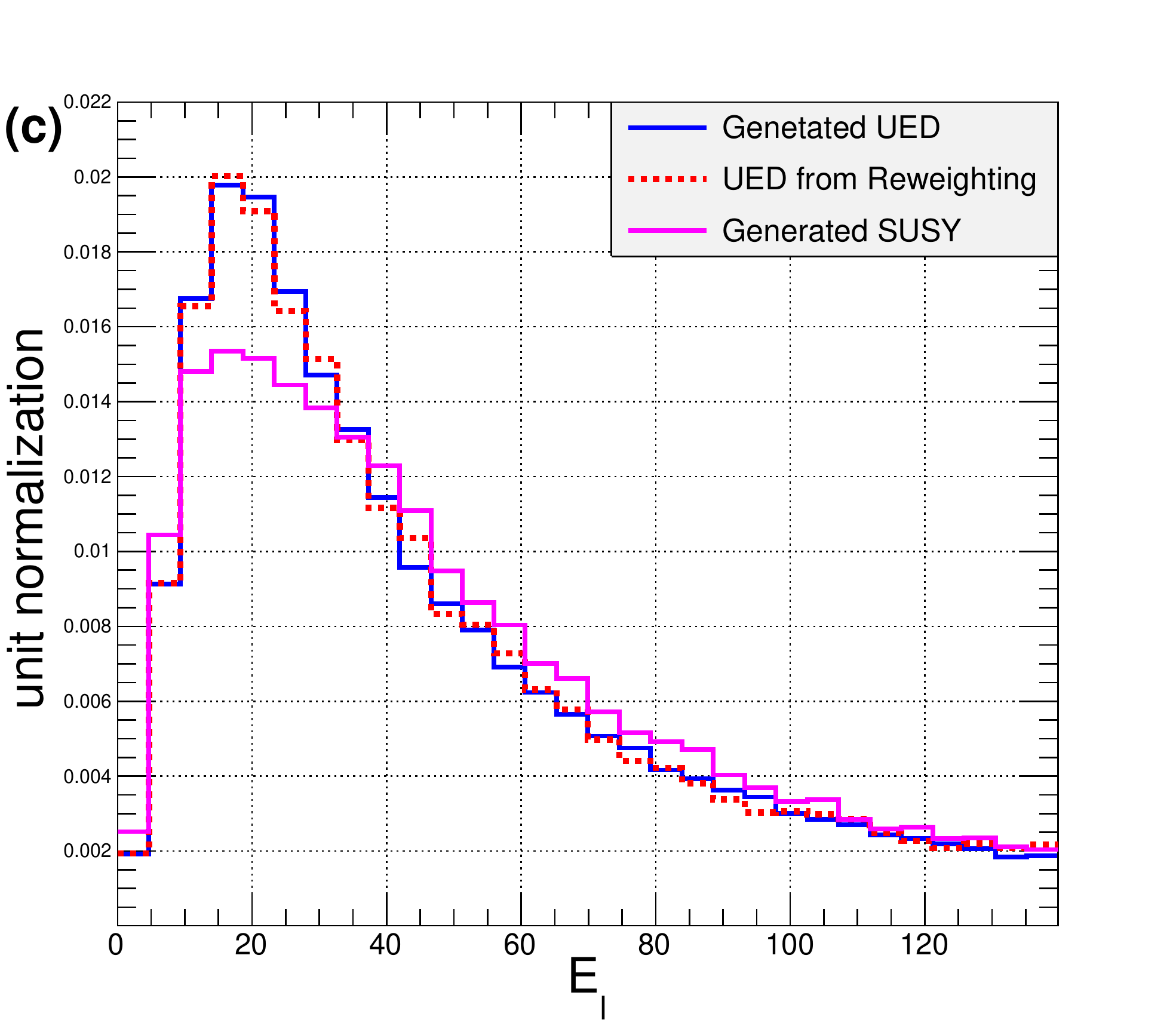} \\
\includegraphics[width=0.328 \textwidth]{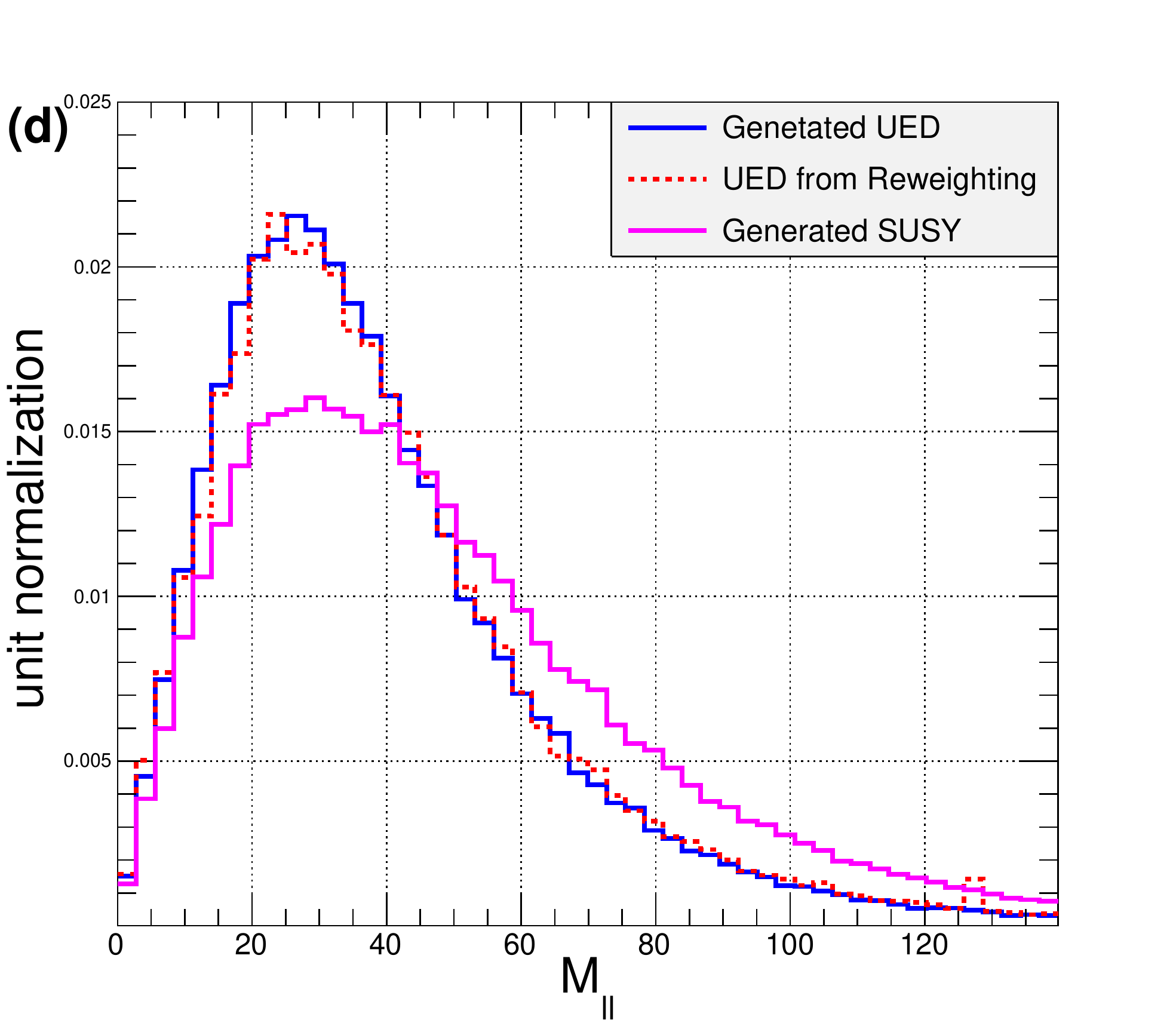}
\includegraphics[width=0.328 \textwidth]{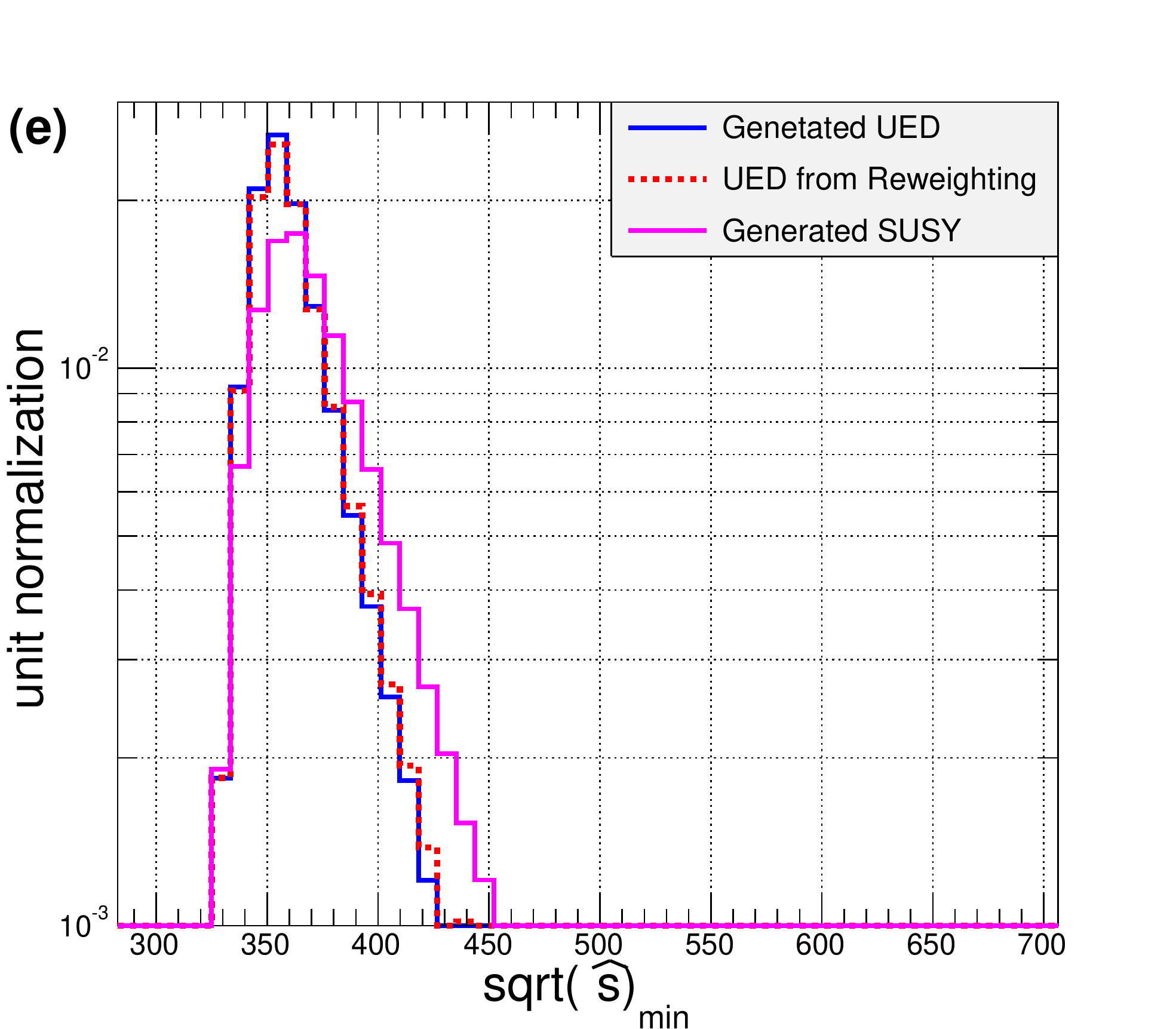}
\includegraphics[width=0.328 \textwidth]{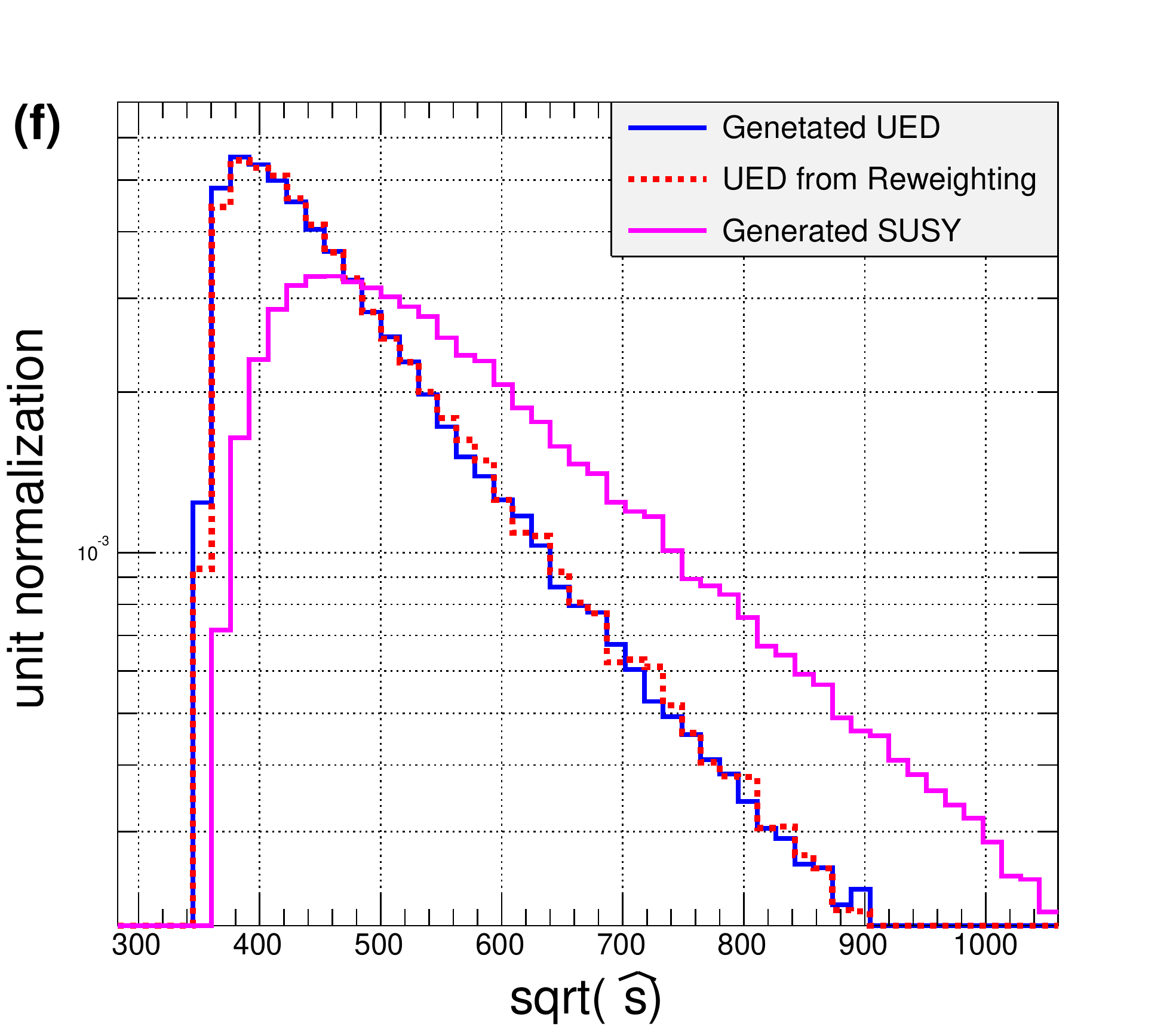}
\end{center}
\vspace{-20 pt}
\caption{
  Distributions of various parton-level quantities which might be used to distinguish
  SUSY from MUED~\cite{Edelhauser:2012xb}.
  The SUSY parameter point used is the CMS LM6 study point \cite{LM6}; 
  in the UED scenario the Kaluza-Klein partners have the same mass as the corresponding superpartners.
 \label{fig:antOBS}}
\end{figure}
%%%%%%%%%%%%% End OF FIGURE %%%%%%%%%%%%%%%%%%%%%%%%%%%%%%%%%

\begin{itemize}
\item{$\cos(\theta_B)$:} A variable calculated from the pseudorapidity
  difference between the two leptons~\cite{Barr:2005dz}, defined by
\begin{eqnarray}
\cos(\theta_B) &=& \tanh\left(\frac{\Delta \eta_{12}}{2}\right), \\
\Delta \eta_{12} &=& \eta_{1} -\eta_{2}, \textrm{ with } \eta_i = \frac{1}{2}\ln\left(\frac{P_i+P_{iz}}{P_i -P_{iz}}\right).
\end{eqnarray}
\item{$\Delta \phi$:} The azimuthal angular difference of the visible particles, 
\begin{equation}
\Delta \phi = \cos^{-1}\left(\frac{\vec P_{1T} \cdot \vec P_{2T}} {|\vec P_{1T}|\,|\vec P_{2T}|} \right).
\end{equation}
\item{$E_\ell$:} The energy of a lepton $E_\ell.$ 
\item{$M_{\ell^-\ell^+}$:} The invariant mass of the two leptons.
\item{$\sqrt{\hat s}_{\min}$:} An estimator of the mass scale of the
  hard scattering, proposed in~\cite{Konar:2008ei,Konar:2010ma},
  calculated for the antler event topology assuming a particular value
  for the $\tilde \chi_1^0$ mass.  Since in practice we would scan
  over this ``trial'' mass, for definiteness we use the correct or ``truth'' value of
  this quantity in the analyses presented here.  The definition of
  $\sqrt{\hat s}_{\min}$ is  
\begin{equation}
\sqrt{\hat s}_{\min} = \sqrt{M_{\ell^-\ell^+}^2+|\vec P_{1T}+\vec
  P_{2T}|^2}+\sqrt{4 m_{\tilde \chi_1^0}^2+\met^2}\, .
\end{equation}
\item{$\sqrt{\hat s}$:} The actual partonic center-of-mass energy.
\end{itemize}
Distributions of the above observables are shown in
Fig.~\ref{fig:antOBS}.  For the various quantities, we show 
\begin{enumerate}
\item (solid magenta lines) The distribution of the quantity obtained from MC simulation of
  the CMS LM6~\cite{LM6} SUSY parameter point, where the mass of the
  right-handed scalar lepton, $m_{\tilde \ell_R}$ is $176.62$ GeV and
  the neutralino mass $m_{\tilde \chi_1^0}$ is $158.18$ GeV.
  \item (solid blue lines) The distribution of the quantity obtained from MC simulation of a
  MUED parameter point where the mass of the Kaluza-Klein (KK) lepton
  is $176.62$ GeV, and the mass of the KK photon is $158.18$ GeV,
  matching the masses of the SUSY particles in the LM6 scenario. 
  \item (dotted red lines) The distribution of the quantity that one would obtain for the
  MUED parameter point, evaluated by the reweighting of events generated
  for the SUSY parameter point.
\end{enumerate}
The MC samples used to generate these distributions consist of
$100,000$ parton level events at the $14$ TeV LHC. 
Fig.~\ref{fig:antOBS} suggests that MUED analyses can in fact be
performed by reweighting SUSY MC events, allowing the efficient
probing of non-standard spin assignments at the LHC.
One could extend this technique beyond MUED to all other possible spin
assignments for a given event topology with given masses.  
In particular, one could set robust and model independent limits on new physics 
in the following way. Among all possible spin assignments 
(or more generally, parameter values), identify the case which
is most difficult to test experimentally, and use it to present limits from
experimental searches. The conservative bounds set in this way 
will be valid for all other spin assignments as well.

\subsection{Jet Clustering, Event Selection, and Detector Response}
\label{sec:clustering}

%%%%%%%%%%%%% Beginning OF FIGURE ################%%%%%%%%%%%%%
\begin{figure}[t!]
\begin{center}
\includegraphics[width=0.8\textwidth]{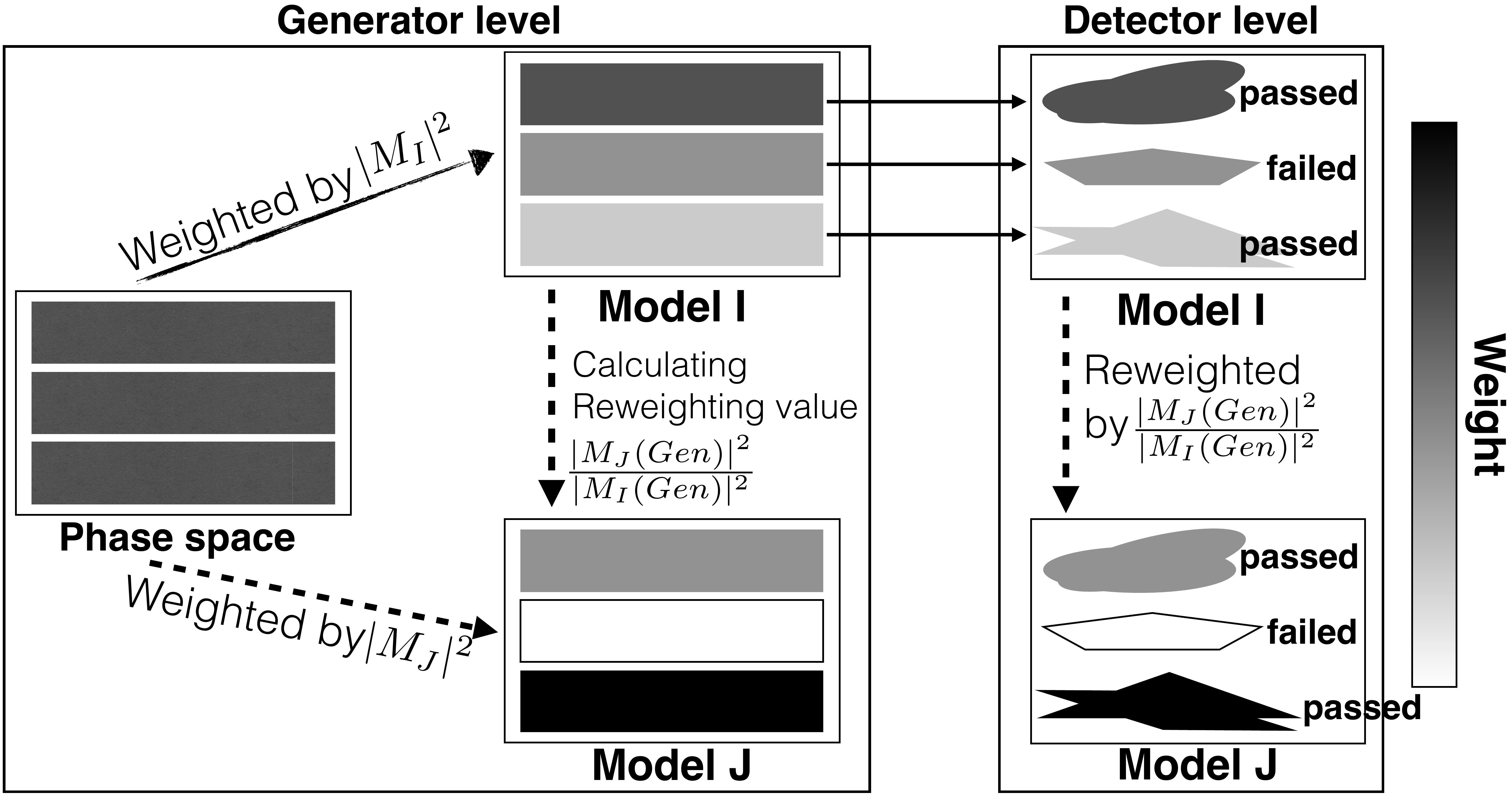} 
\end{center}
\vspace{-20 pt}
\caption{A conceptual diagram illustrating the use of reweighting 
  for events which are passed through detector simulation. 
  As noted above, if a physics model, $I$, shares the same phase space
  with a model $J$,  we can perform Monte Carlo analyses of model $J$ by
  reweighting detector level events of model $I$ using a reweighting value
  calculated using ``truth'' MC events at parton or generator level,
  $\frac{|M_J(Gen)|^2}{|M_I(Gen)|^2}.$  We denote actual event
  generations by solid lines and mark calculations for the reweighting
  procedure with dashed lines. 
  The weight of an event in the given model is indicated by the shade
  of gray.  The effects of showering, hadronization, detector
  simulation, and jet clustering are shown conceptually by the
  reshaping of the events which are rectangular at generator level and
  oddly shaped after detector simulation.  Clearly the detector level
  events are the same whether or not we are reweighting (hence have
  the same shapes in our figure).  Also, detector level events either
  pass cuts or fail cuts regardless of whether reweighting is
  employed.  
\label{fig:detector}}
\end{figure}
%%%%%%%%%%%%% End OF FIGURE %%%%%%%%%%%%%%%%%%%%%%%%%%%%%%%%%

Parton level information can be distorted by various factors,
including showering, hadronization, jet clustering, and detector
response.  All of these can ``deform'' the phase space of generated
particles. But, as noted in Section~\ref{sec:mc},
the relative weights in various models for a given
observed event depend only on the parton level amplitudes
for the event in each model.  We present a conceptual picture of this
reweighting procedure in Fig.~\ref{fig:detector}.

\begin{figure}[t]
\begin{center}
\includegraphics[width=0.95\textwidth]{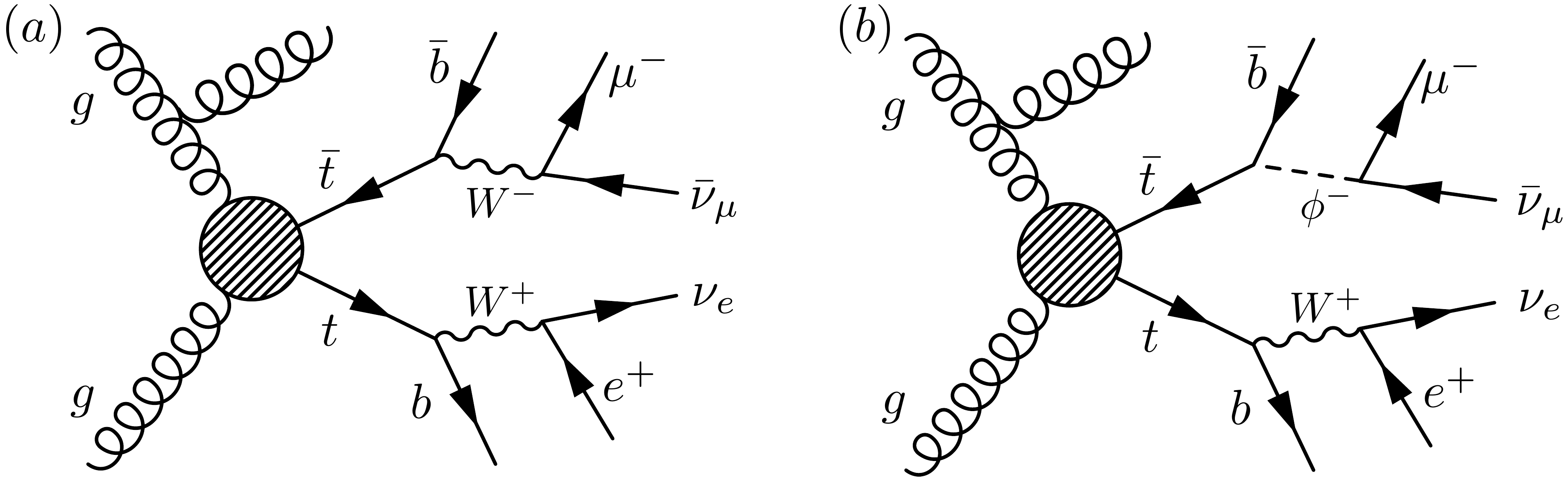} 
\end{center}
\vspace{-20 pt}
\caption{To demonstrate the use of reweighting events involving
  detector simulation and jet clustering, we study top quark pair
  production, followed by decays involving a $b$-jet and either (a) the
  $W^{\pm}$ boson, as in
  the SM, or (b) a charged Higgs, $\phi^{\pm}$.  
  As we show, the charged Higgs scenario (b) can be analyzed using
  SM dilepton $t\bar t$ events (a) through reweighting.  Due to the
  mass dependence of Higgs couplings, the lepton from charged Higgs
  decay is always a muon and never an electron.\label{fig:ttbar}}
\end{figure}

To illustrate this point numerically, we simulate $t \bar t$ production at
$14$ TeV  LHC, where the top-quark (anti-top quark) decays into a $W^{+}$ ($W^{-}$) boson
or charged Higgs $\phi^{+}$ ($\phi^{-}$), see Fig.~\ref{fig:ttbar}.
Our goal is to evaluate the performance of the reweighting procedure
described in Fig.~\ref{fig:detector}, including effects that we did not account for
in the parton-level examples described in Sections~\ref{sec:h4l} and \ref{sec:antler}.
Specifically, we generated $100,000$ parton level events with
{\tt MadGraph5(aMC)} v.2.1.0, passed the events to {\tt PYTHIA} 6.4,
and performed detector simulation with {\tt Delphes} v.3.0.12. For jet
clustering, we used the anti-$k_t$ algorithm with $\Delta R=0.5$. For
the reconstruction of visible particles, we employ default smearing factors
and lepton tagging efficiencies. We apply basic analysis cuts with
$p_T(e) > 10$ GeV, $p_T(\mu) > 5$ GeV and $p_T(j)>30$ GeV and do not
require $b$-tagging in our analysis. 

%%%%%%%%%%%%% Beginning OF FIGURE ################%%%%%%%%%%%%%
%%%%%%%%%%%%% End OF FIGURE %%%%%%%%%%%%%%%%%%%%%%%%%%%%%%%%%

\begin{figure}[t]
\begin{center}
\includegraphics[width=0.328 \textwidth]{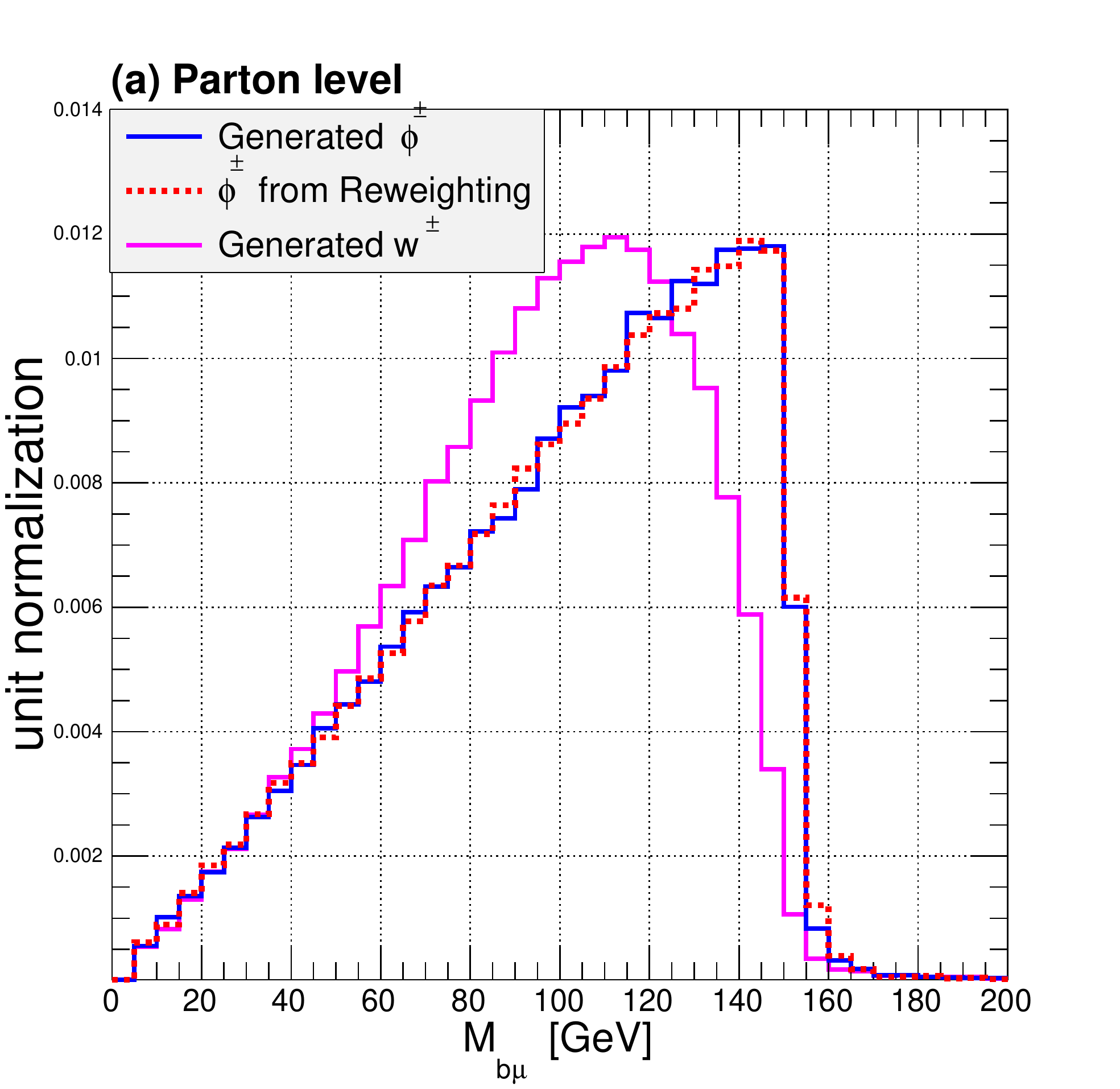}
\includegraphics[width=0.328 \textwidth]{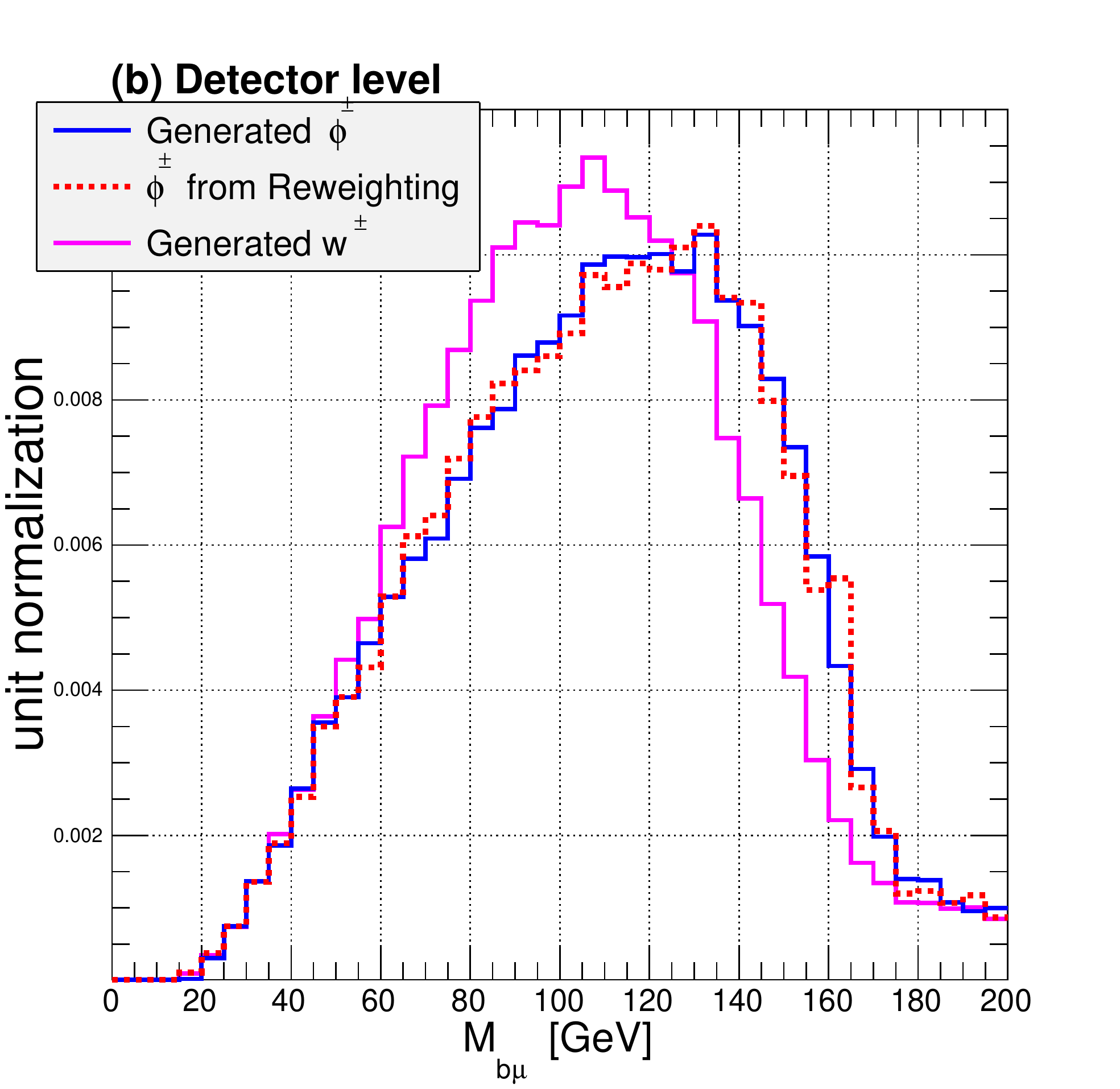}
\includegraphics[width=0.328 \textwidth]{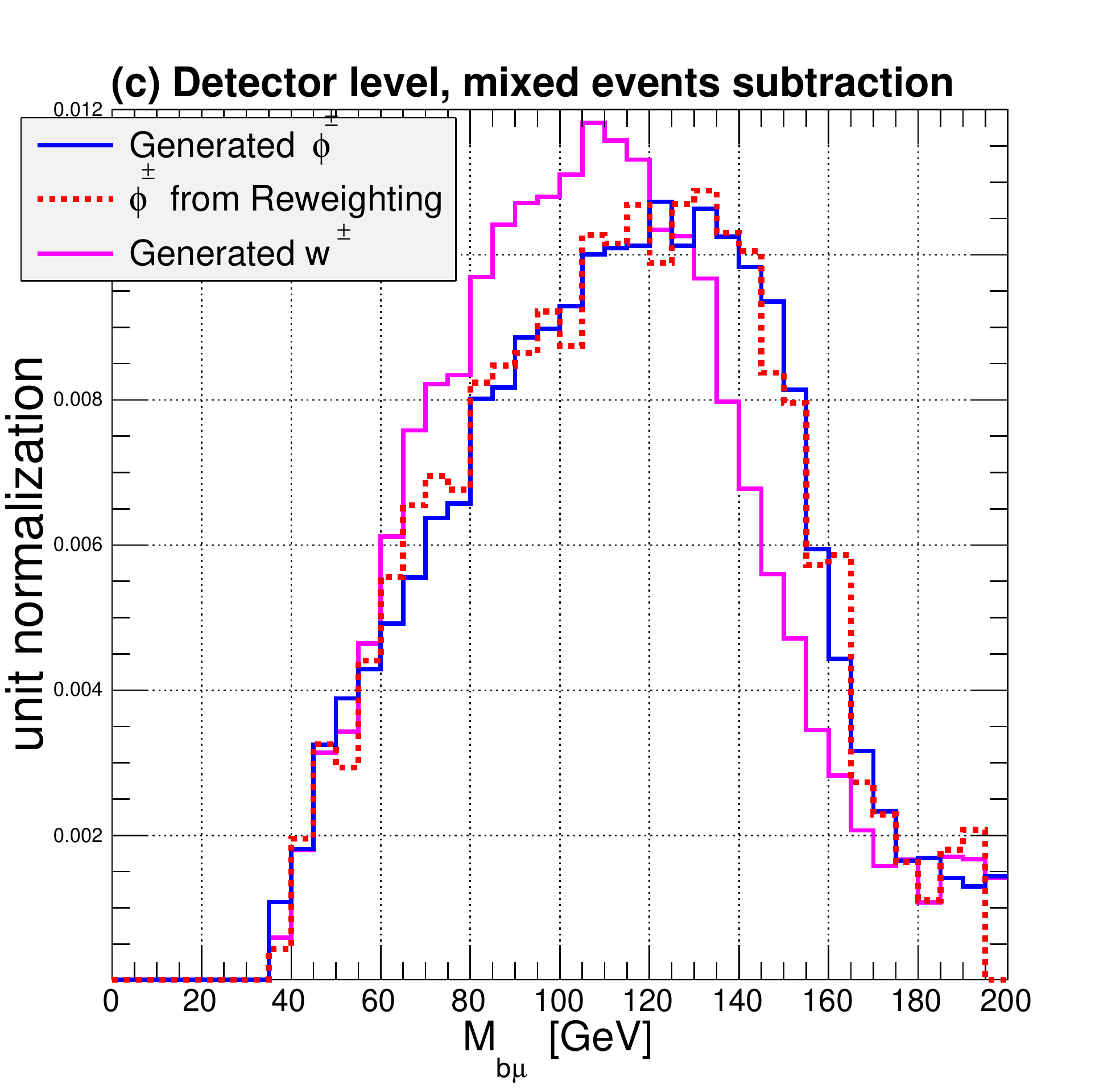}
\end{center}
\vspace{-20 pt}
\caption{Invariant mass distributions, $m_{b\mu}$, of $b$-jets and the
  muon for events generated using the topologies shown in 
  Fig.~\ref{fig:ttbar}.  The variable
  $m_{b\mu}$ contains information about the spin of the intermediate
  particle, $W^{\pm}$ or $\phi^{\pm}$.
  In (a) we plot this quantity at parton level, pairing the muon with
  the $b$ from the same top decay.
  In $(b)$, we use events which have undergone realistic
  simulation, i.e. they are showered, hadronized, and
  passed through fast detector simulation and jet clustering.  
  To separate the effects of detector simulation, etc. from the
  effects of combinatorics, we use truth information to select the
  correct jet to pair with the lepton.
  In $(c)$, which was also generated using realistic simulation, we include
  the effects of combinatorics by not using truth information 
  (to model the actual experimental situation) and
  instead using the mixed event subtraction method~\cite{Hinchliffe:1996iu} to suppress
  contributions from incorrect parings of jets and muons.\label{fig:invDET}}
\end{figure}

When the top quark decays into a bottom quark and a charged Higgs boson, the
invariant mass distribution of the bottom quark and the corresponding
lepton from the charged Higgs boson decay will have a triangular shape
(the solid blue line in Fig.~\ref{fig:invDET}(a)).\footnote{The charged Higgs 
is a scalar and cannot ``communicate'' information about the helicity of 
the $b$-quark to the lepton, thus the shape of the distribution is as 
as in the pure phase space case.} On the other hand, the invariant mass
distribution of the $b$-quark and the corresponding lepton from a $W^\pm$
decay will have a non-trivial shape (solid magenta lines in Fig.~\ref{fig:invDET}(a)). 
These two cases can be easily related by reweighting at the parton level, 
as shown by the red dotted histogram in Fig.~\ref{fig:invDET}(a).

At the detector level, the $b$-quark is reconstructed as a jet, thus
we select the two highest $p_T$ jets to be our $b$ jets.
In order to study the effects of combinatorics separately, in
Fig.~\ref{fig:invDET}(b) we show an intermediate (and unphysical) 
case where we use MC truth information to select the correct jet 
to be paired with the muon. We see that the reweighting procedure correctly
reproduces the distributions of kinematic variables computed in terms 
of reconstructed objects.

Finally, in Fig.~\ref{fig:invDET}(c) we show the same three distributions, 
only this time ignoring the MC truth information and fully accounting for 
the combinatorics. In order to reduce the
contamination from combinatorial pairing errors, we use the mixed
event subtraction method~\cite{Hinchliffe:1996iu}, where we add
both possible pairings in a given event, then subtract a pairing of the muon with a 
jet from a different event.
We emphasize that the point of these analyses is that a distribution made
through reweighting is identical to the actual distribution made with
a MC events generated for the hypothesis in question.

%%%%%%%%%%%%% Beginning OF FIGURE ################%%%%%%%%%%%%%
%%%%%%%%%%%%% End OF FIGURE %%%%%%%%%%%%%%%%%%%%%%%%%%%%%%%%%

\section{Possible extensions}
\label{sec:practical}

As noted in Section~\ref{sec:applications} above, reweighting events
generated using model $I$ to obtain distributions in model $J$ may become
problematic for two main reasons. Firstly,  
in a phase space region where the cross
section for the new model $J$ is high, while the 
cross-section for the reference model $I$ is low,
the region is sampled by relatively few MC events.
Secondly, the spread in the reweighting factors for events in the given
bin may become large. Both of these effects are described by
Eq.~\ref{eq:weighted-error}.  We are obviously interested in
minimizing the errors on histograms, but at the same time, we want to keep 
as low as possible the number of events which need to be processed 
by full MC simulation.  

We therefore suggest three practical procedures that can be used to 
mitigate issues related both 
to undersampling and to the spread on reweighting factors,
and possibly to help understand systematic effects.

\begin{enumerate}
\item {\bf Targeted Generation:}  In this procedure, whenever the
  error in a reweighted histogram obtained for model $J$ 
  becomes greater than a certain threshold, one generates additional new events
  in the undersampled region that are phase space ``neighbors'' of the events
  already in the bin. The
  simplest way to think of this is to divide phase space $\mathcal{P}$
  into two non-overlapping regions, $\mathcal{N}$ and
  $\overline{\mathcal{N}}$, where $\mathcal{N}$ describes a phase space
  region where additional event generation is desired, and
  $\overline{\mathcal{N}}$ is the rest of the phase space.
  Then we generate events for model $J$ only in region $\mathcal{N}$.
  Distributions for $J$ will then be obtained by reweighting the events
  generated for model $I$ in the $\overline{\mathcal{N}}$ and for events
  generated for model $J$ in the region $\mathcal{N}$.\footnote{The two samples 
  must be mixed accounting for the relative number
  of events generated in regions $\mathcal{N}$ and
  $\overline{\mathcal{N}}$.   This is an example of
  multi-channel integration with channels $I$ and $J$.}
If one then
wishes to obtain distributions for a third model, $K$, one will
reweight events generated for model $I$ in region
  $\overline{\mathcal{N}}$ and the events generated for model $J$ in
  region $\mathcal{N}$, in each case using a reweighting factor
  involving the appropriate models.
  
\item {\bf Regions of Overlap:}
  A related approach is to generate large samples of unweighted events at
  several benchmarks in parameter space, which we label
  $I_1$, $I_2$, etc.  One then obtains distributions $D_1$, $D_2$,
  etc. for model $J$ by reweighting events generated for model $I_1$,
  $I_2$, etc.\footnote{This is another example of multi-channel integration.} 
  If the benchmarks are chosen appropriately, every
  important region in phase space for model $J$ 
  can be described with low errors by reweighted events generated for
  some model $I_j$, hence avoiding the undersampling issues described above.
  Also, the comparison of the values of the distribution obtained from
  different choices of unweighted events may allow
  for the calculation of systematic errors from this procedure.

\item {\bf Optimized Sample:}
A third, and quite distinct approach to optimizing reweighting
procedures for large parameter space is to optimize the choice of
model $I$ to avoid undersampling in as much of the parameter space as
possible.  In general, this would be performed by maximizing an
integral over the parameter space (possibly weighted by some prior or
importance measure) of a function representing the 
errors that will be obtained on bins of the desired histograms.
This appears to be a hard integral to perform in practice, so reasonable
approaches will probably involve approximation to some degree. 
It is also unclear whether even an optimized model
  would be as efficient as generating events for at least the several
  benchmarks in parameter space described in the previous item.
\end{enumerate}

\section{Conclusions}
\label{sec:conclusions}

We have presented and described in detail a procedure aimed at surmounting
the experimental challenges presented by the multiplicity of
theoretical models, which may each, in turn, have large parameter spaces.  
We realize that the procedure presented here is not totally unknown to
experimentalists (particularly in the special case of parton
distribution function reweighting~\cite{Tricoli:2008ev}).  
However, given the importance of this challenge,
we felt it important to highlight the potential of this method,
especially for the study of large BSM parameter spaces.

Specifically, we have shown how reweighting events using ``truth''
information from generator-level Monte Carlo events, including the
momenta of invisible particles, makes possible the detailed study of
large signal parameter spaces and/or large numbers of signal models at
the level of detail needed for experimental purposes.  We demonstrated
this in several motivated physics examples and also illustrated
potential issues which can arise, 
including an explicit discussion of the errors on the 
weighted histograms generated from reweighting.

There are several advantages of our method:
\begin{itemize}
\item Given a large existing sample for some theory model $I$, 
we gain speed in generating an effective fullsim sample for another theory model $J$
by avoiding detector simulation, hadronization, showering and fragmentation.
\item All existing fullsim samples can be resurrected for the study of new models.
\item The reweighting is computationally very quick and simple - for example,
there is no need to integrate over the momenta of any invisible particles.
\item Reweighting also allows the interactions between theorists and
  experimentalists to be maximally efficient -
the experimentalists are handling the fullsim MC generation, while the 
theorists are providing only the parton-level functions for reweighting the events.
\end{itemize}

We conclude with the simple recommendation that in order to be able to implement 
the procedure described here, when generating fullsim samples, one should make 
sure to record and store the generator-level information which is needed for reweighting.

\acknowledgments

JG, JL, KM, and SM thank their CMS colleagues for useful discussions,
and in particular, G. Hamel de Monchenault for useful comments on a 
draft of this manuscript.  
MP appreciates a useful discussion with J. Wacker on the application
of a matrix element reweighting method for simplified model searches
 at the ``Coordinating a Simplified Models Effort'' workshop,  CERN
 2013. All authors thank the anonymous referee for useful comments,
especially on the treatment of uncertainties in weighted histograms
and on using this technique to obtain model-independent limits.
MP is supported by the World Premier International Research Center
Initiative (WPI Initiative), MEXT, Japan. 
Work supported in part by U.S. Department of
Energy Grant ER41990.  Fermilab is operated by the Fermi
Research Alliance under contract DE-AC02-07CH11359 with the
U.S. Department of Energy.

\end{document}